\documentclass[useAMS,usenatbib]{mn2e}
\usepackage{graphicx}
\usepackage{hyperref}
\hypersetup{
     colorlinks   = true,
     citecolor    = blue,
     linkcolor     = blue
}
\usepackage{caption}
\newcommand{\R}{\mathrm}
\newcommand{\I}{\textit}

\def\apj{ApJ}
\def\apjs{ApJS}

\def\aap{A\&A}
\def\aaps{A\&AS}
\def\aj{AJ}
\def\mnras{MNRAS}

\def\pasp{PASP}
\def\nat{Nature}

\def\araa{Ann. Rev. Astron. Astrophys.}

\title[The H {\sc i} content of metal-poor blue compact dwarf galaxies]{The H {\sc i} content of extremely metal-deficient blue compact dwarf galaxies}
\author[T. X. Thuan et al.]{T.~X.~Thuan$^{1}$
, K.~M.~Goehring$^{1}$, J.~E.~Hibbard$^2$, Y.~I.~Izotov$^3$ and L.~K.~Hunt$^4$ 
\\
$^{1}$Astronomy Department, University of Virginia, P.O. Box 400325, Charlottesville, VA 22904-4325, USA; txt@virginia.edu, \\ kmg4mx@virginia.edu \\
$^{2}$National Radio Astronomy Observatory, Charlottesville, VA 22903, USA;
jhibbard@nrao.edu\\
$^3$Main Astronomical Observatory, National Academy of Sciences of Ukraine, 03680 Kyiv, Ukraine; izotov@mao.kiev.ua\\
$^4$INAF-Osservatorio Astrofisico di Arcetri, Largo E. Fermi 5, I-50125 Firenze, Italy; hunt@arcetri.astro.it} 

\begin{document}

\date{19 January 2015}


\maketitle


\begin{abstract}
 We have obtained new H~{\sc i} observations 
with the 100 m Green Bank Telescope (GBT)  
for a sample of 29 extremely metal-deficient 
star-forming Blue Compact Dwarf (BCD) galaxies,
selected from the Sloan Digital Sky Survey spectral data base to be 
extremely metal-deficient (12 + log O/H $\leq$ 7.6). 
Neutral hydrogen was detected in 28 galaxies, a 97\% detection rate. 
Combining the H~{\sc i} data with SDSS optical spectra for the BCD sample 
and adding complementary galaxy samples from the literature to extend the 
metallicity and mass ranges, we have studied  how the H~{\sc i} content of a galaxy 
varies with various global galaxian properties.
There is a clear trend of increasing gas mass fraction with decreasing 
metallicity, mass and luminosity.  
We obtain the relation $M$(H~{\sc i})/$L_g$
$\propto$ $L_g^{-0.3}$, in agreement with 
previous studies based on samples with a smaller luminosity range. 
The median gas mass fraction $f_{gas}$ for the GBT sample 
is equal to 0.94 while the mean gas 
mass fraction is 0.90$\pm$0.15, with a lower limit of $\sim$0.65. 
The H~{\sc i} depletion time 
is independent of metallicity, with a large scatter around the 
median value of 3.4 Gyr. The ratio of the baryonic mass to the dynamical mass
of the metal-deficient BCDs 
 varies from 0.05 to 0.80, with a median value of $\sim$0.2. About 65\% 
of the BCDs in our sample have an effective yield larger than the true yield,
implying that the neutral gas envelope in BCDs 
is more metal-deficient by a factor of 1.5--20, as compared to the ionized gas.  


\end{abstract}


\section{Introduction}\label{intro}

The formation and evolution of the first galaxies in the universe remains a key issue in cosmology. It is now thought that large massive 
star-forming galaxies form in a hierarchical manner from the assembly of smaller dwarf systems, through accretion and merger processes. 
These galaxy interactions 
 trigger the formation of stars which enrich the interstellar gas in metals 
by stellar winds and supernovae. In this scenario, the number of extremely metal-deficient (XMD) dwarf galaxies should be high in the early universe but considerably smaller at the present epoch \citep{M12}.   
  
However, while much progress has been made in finding large populations of galaxies at 
high ($z\geq$3) redshifts \citep[e.g. ][]{Ste03,Ade05}, truly 
chemically unevolved galaxies 
remain elusive in the 
high-$z$ universe. The spectra of distant galaxies generally indicate the presence of a substantial amount of heavy elements, implying previous star formation and metal enrichment. The discovery of XMD galaxies at high $z$ may have to wait until the advent of the 
{\sl JWST} and 30 m-class ground-based telescopes.

We adopt here a different approach. Instead of searching for 
high-$z$ metal-deficient objects, we focus our attention on 
XMD star-forming dwarf galaxies in the local universe. They are the most promising local proxies of chemically unevolved galaxies in the early universe, 
and are usually found among a class 
of dwarf galaxies undergoing intense bursts of star formation called Blue Compact Dwarf (BCD) galaxies \citep{Thu81}. The optical spectra of the 
BCDs are characterized by a blue continuum on which are superimposed 
strong narrow emission lines. XMD BCDs are very rare \citep*{Izo12}. For more than three decades, one of the first BCD discovered, I~Zw~18 \citep{Sar70}, held the 
record as the most metal-deficient emission-line galaxy known, with an oxygen abundance [O/H]= 12 + log O/H = 7.17$\pm$0.01 
in its northwestern component and 7.22$\pm$0.02 in its southeastern component \citep{TI05} \citep[$\sim$ 3\% solar, adopting the solar abundance 12+logO/H = 
8.76 of ][]{Ste15}. Only in 2005 has I~Zw~18 been superceded in its rank 
by SBS~0335--052W with a metallicity [O/H] = 7.12 \citep*{ITG05}.

Because of the scarcity of XMD emission-line galaxies, we stand a much 
better chance of discovering them in very large spectroscopic surveys such as 
the Sloan Digital Sky Survey \citep[SDSS, ][]{Y00}. We have carried out 
a systematic search for such 
objects with [O/H] in the SDSS spectroscopic data release 
7 (DR 7) \citep{Aba09}.  Imposing cut-offs in metallicity and redshift  
results in a total sample of 29 XMD BCDs. Similar searches for 
XMD galaxies in the SDSS have been carried out by \citet{Mor11} and 
\citet{San16}. We found a total of 10 galaxies in common between 
the present sample and that of \citet{San16} so that the two samples are likely to 
have similar properties, and the H~{\sc i} characteristics discussed here probably apply to the
\citet{San16} objects as well.   

The focus of this paper is the study of the 
neutral hydrogen content of these XMD objects. There have 
been previous H~{\sc i} studies of this type of extremely metal-poor galaxies. 
Thus, \citet{Fil13} have carried out 
a single-dish H~{\sc i} study with the Effelsberg radio telescope 
of a subsample of 29 XMD galaxies selected from the \citet{Mor11} list.    
We will use part of the data of those authors to supplement our own and will compare with our results with theirs when warranted. A handful of interferometric  H~{\sc i} maps of other XMD galaxies have also been obtained by the Pune group 
with the 
Giant Metrewave radio telescope \citep*[see ][ and references therein]{Ekt09,Ekt10}.     
 In Section 2, we define the XMD BCD sample and describe observations 
of their neutral hydrogen content with the Robert C. Byrd Green Bank Telescope (GBT) at the National Radio Astronomy Observatory~\footnote{The National Radio Astronomy Observatory is a facility of the National
Science Foundation operated under cooperative agreement by Associated
Universities, Inc.}. This
sample will be referred to hereafter as the GBT sample. 
Section 3 describes the H {\sc i} data along with  
derived ancillary data such as metallicities, star formation rates, stellar masses needed to study trends of the neutral gas content with other properties 
of the 
BCDs. In Section 4, we discuss several comparison samples compiled from the 
literature, useful for      
studying the H~{\sc i} content of star-forming 
galaxies over a wider range of heavy element abundances and stellar masses. 
Our total galaxy sample, composed of the GBT and three comparison samples,    
includes 151 objects and covers the extensive metallicity 
range 7.20 $\leq$ [O/H] $\leq$ 8.76.
We study in Section 5 the correlations of various quantities. 
In particular, we analyze 
the dependence of the neutral gas mass to light ratio on 
metal abundance, and that of the gas mass fraction on stellar mass.
We also discuss the chemical evolution of XMD BCDs. 
We summarize our conclusions in Section 6. Throughout this paper,
we adopt the 
cosmological model caracterized by 
a Hubble constant $H_{0}$ = 73 km s$^{-1}$ Mpc$^{-1}$, 
a matter density parameter 
$\Omega_M$ = 0.27 
and a dark energy density parameter 
$\Omega_\Lambda$ = 0.73 \citep{R11}.

\section{H {\sc i} observations}

\subsection{The GBT sample}

We have constructed our XMD BCD sample by 
applying the following selection criteria to the SDSS DR7 spectral database. As 
described in \citet{Izo12}, we first select them on the basis of 
the relative fluxes of particular emission lines as measured on the SDSS 
spectra: [O~{\sc iii}]$\lambda$4959/H$\beta$ $\leq$ 1
and [N~{\sc ii}]$\lambda$6853/H$\beta$ $\leq$ 0.1.  
These spectral properties select out uniquely 
low-metallicity dwarfs since no other type of galaxy possesses them.
 Second, since one of the main scientific objectives here  
is to study how the H~{\sc i} properties of star-forming galaxies vary 
with metallicity, it is important to have accurate heavy element abundances 
for the XMD BCDs.
Thus, we have included in our sample 
 only those BCDs that have a well-detected [O~{\sc iii}]$\lambda$4363 
line as this electron temperature-sensitive emission line allows a direct 
and precise abundance determination. 
Third, we have set a metallicity cut-off [O/H] 
$\leq$ 7.6 ($\sim$8\% solar) to choose only XMD galaxies. 
This metallicity threshold  
has been suggested by \citet{Izo99} to characterize  
very young galaxies, with most  
of their stellar populations formed not more than $\sim$1 Gyr ago. 
Lastly, we have chosen the BCDs to be not more distant than $\sim$85 Mpc, 
so we can measure their H {\sc i} emission with a good signal-to-noise ratio 
within a reasonable 
integration time with the GBT.
This distance upper 
limit corresponds to a recession velocity of $\sim$6200 km s$^{-1}$.   
These selection criteria 
result in a total sample of 29 XMD BCDs. In contrast to many 
previous BCD samples that have been observed in H {\sc i} 
\citep*[e.g. ][]{Thu81,Thu99,Huc05,Fil13}, 
this BCD sample 
is unique in that : 1) it contains only very metal-deficient objects 
(7.35 $\leq$ [O/H] $\leq$ 7.60) as compared to previous samples which include  
many objects with [O/H]$\geq$ 7.7. This allows us to study very chemically unevolved galaxies; 2) it has O abundances determined precisely for each object,
using the direct method based on  
the [O~{\sc iii}]$\lambda$4363 line. This is in contrast 
to the approximate abundances 
derived for many galaxies in previous samples, using the 
statistical strong-line method, because of the undetectability 
of the [O~{\sc iii}] line; 3) all objects possess accurate SDSS images and 
photometry with which we can derive optical luminosities and mass-to-light 
ratios.    
 
\subsection{Observations and Data reduction}

The observations were obtained with the 100 m GBT during the periods September 2005 and 
January-February 2006. The GBT spectrometer backend was used with a total bandwidth of 12.5 MHz and 9-level sampling,
resulting in a total of 16,384 spectral channels. We use 
two spectral windows centered at the 
same frequency (1420.4058 MHz) with two different linear polarizations 
(XX and YY).
Each target was observed in total power mode 
with multiple (between 2 and 6) 10 minute on-source, 10 minute off-source 
pairs. The total on-source time for each galaxy was determined during the
observing runs, depending on the strength of the spectral 
feature relative to the noise. 
Since the two polarizations were 
detected independently, they were averaged to improve sensitivity.   
The data were flagged for radio frequency interference (RFI) and all scans summed. The data were calibrated by observations of standard continuum calibrators 
from the list of 
NRAO VLA Sky Survey unconfused calibrator sources prepared by J. Condon and Q.F. Yin, using the same backend setup. We checked our calibration by observing standard spectral line calibrators from the list of \citet{Hog07} and found 
it to be good to $\sim$3\%. All data reduction was done in GBTIDL. The data were boxcar-smoothed by 20 channels, for a final frequency resolution of 15.26 kHz, 
corresponding to a velocity resolution of $\sim$3.3 km/s. RMS noise levels were ~3 mJy for a single 10 minute
 on/off scan. Spectral baselines were fit using line-free channels and 
subtracted out. Generally, a low-order fit, between 1 and 3, was used. 
Line fluxes and linewidth at 20\% 
and 50\% of peak intensity 
were measured in IDL. Distances were calculated from the 
measured redshifts, after correcting for Virgo infall, 
using the Virgocentric flow model of \citet{Mou00}.

Of the 29 XMD BCDs in our sample, 28 galaxies were detected. This corresponds 
to a 97\% detection 
rate, considerably higher than those of 
previous BCD samples.
For comparison, the detection rate of the \citet{Thu81} sample is 80\%, 
that of the \citet{Thu99} sample 74\%, and that of \citet{Fil13} only 
34\%. 
The reason for the high detection rate of our GBT sample is probably the higher 
sensitivity of the present observations. Thus, 
the 5$\sigma$ uncertainties of our observations 
are $\sim$0.1 Jy km s$^{-1}$ while those of the \citet{Fil13} observations 
are about 0.6 Jy km s$^{-1}$.

\subsection{H {\sc i} data}



\begin{figure*}
\centering
\hbox{
\includegraphics[scale=1.3,angle=270]{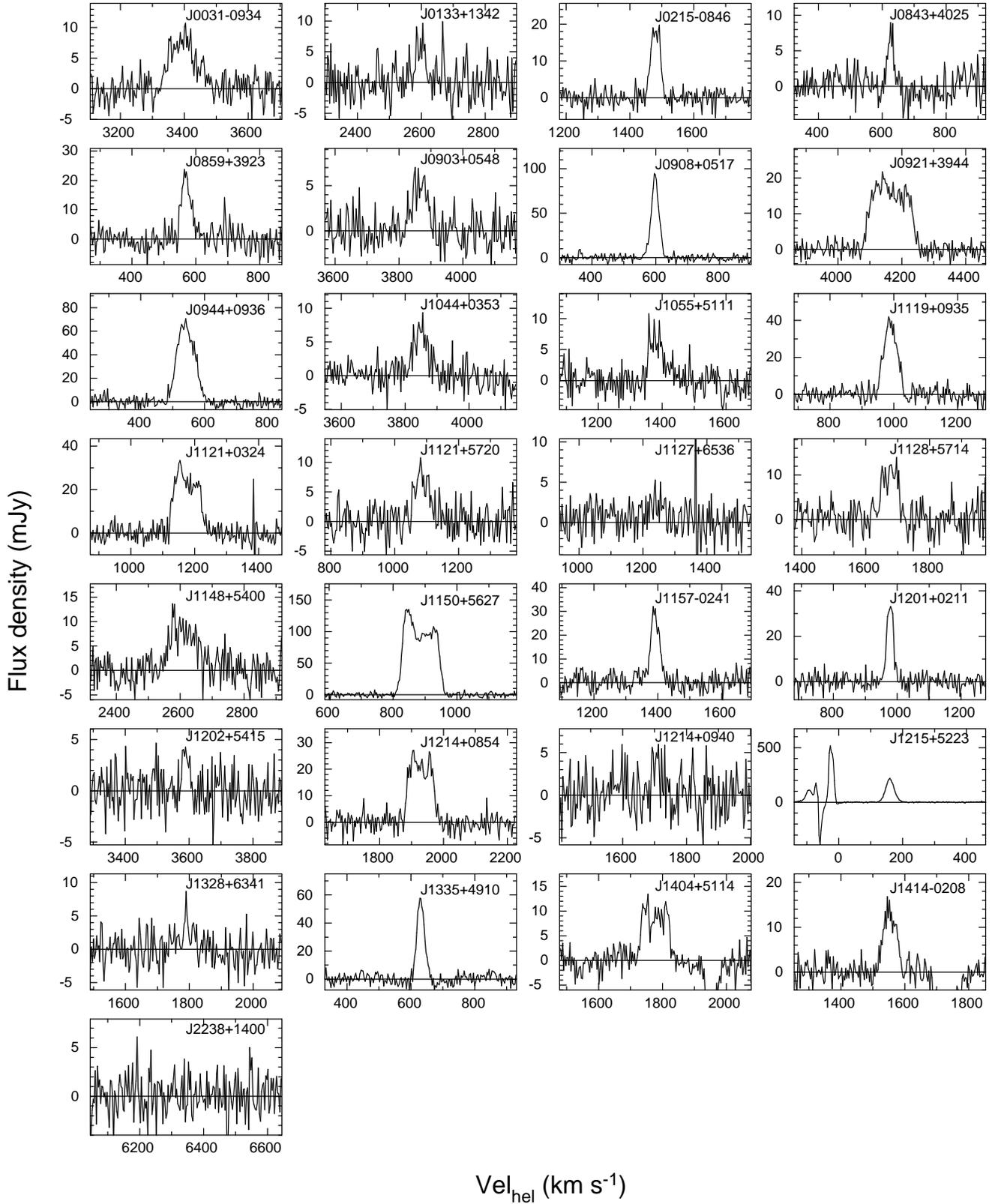}}
\caption{H {\sc I} profiles of the 29 galaxies observed with the GBT. Profiles are arranged, from top left to bottom right, in order of increasing right ascension.}\label{fig1}
\end{figure*}

H {\sc i} profiles (after boxcar smoothing and baseline removal) of all galaxies in the sample are presented in Fig. \ref{fig1}. 
 There is only one non-detection, the BCD J2238+1400 = 
HS~2236+1344. The profiles are 
arranged from left to right and from top to bottom   
in order of increasing right ascension. 
In addition, Fig. \ref{fig2} 
 shows the SDSS image of each galaxy, taken  
from Data Release 10 \citep[DR10, ][]{Ahn13}.
The images are useful for examining the morphology and colors of the 
sample objects and for evaluating potential 
H {\sc i} contamination from neighboring galaxies.  
It is clear that our objects are generally very compact, 
the average half-light $g$ radius of the dwarf galaxies 
in our sample is about 10$\arcsec$ (column 5 of Table \ref{tab2}), 
their linear diameters being generally less than 2 kpc.  Adopting an average 
H {\sc i}-to-optical size ratio of $\sim$3 \citep{Thu81}, the H {\sc i} 
angular extent
of all sample galaxies are considerably smaller than the 9$\arcmin$ GBT beam 
at 21 cm, 
so we did not need to apply any beam correction to the flux densities. 
To check for potential contamination of the H {\sc i} emission   
from neighboring galaxies, we have also examined the SDSS DR10 
images in a square field of 12$\arcmin$ on a side centered on each object.
 We have found no case of contamination.
For J1214+0940 with $v$ = 1702 km s$^{-1}$, there is a faint 
diffuse yellow galaxy at 4\farcm5 of the BCD in the NW direction, 
but its velocity is 1245 km s$^{-1}$.  

\begin{figure*}

\centering

\includegraphics[scale=0.8]{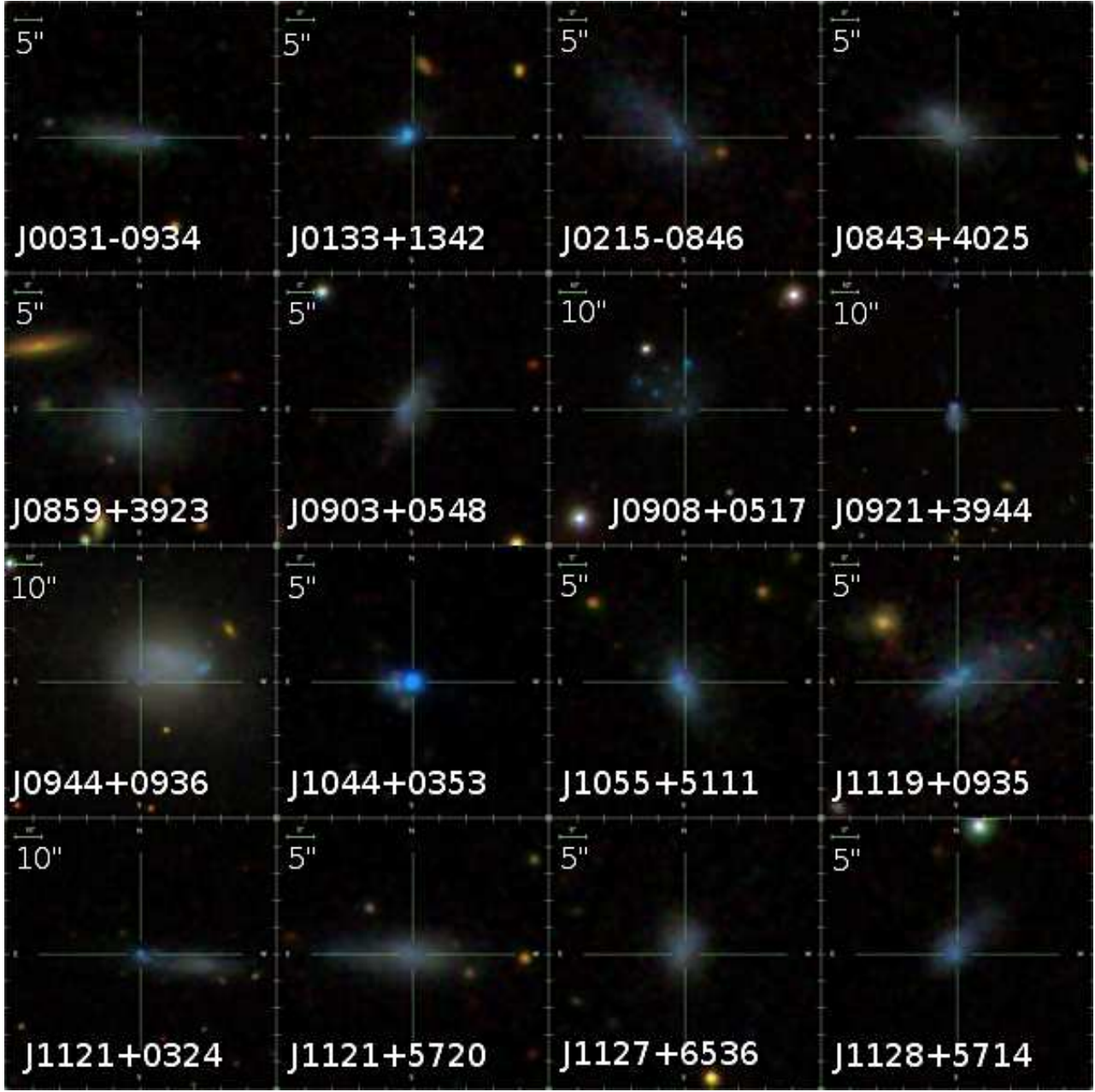}
\caption{SDSS images of the BCDs whose H {\sc i} profiles are shown in Fig.\ref{fig1}. 
North is on top and East to the left. The scale is shown by a horizontal bar.}\label{fig2}
\end{figure*}

\begin{figure*}

\centering

\renewcommand{\thefigure}{\arabic{figure}} 
\addtocounter{figure}{-1}

\includegraphics[scale=0.8]{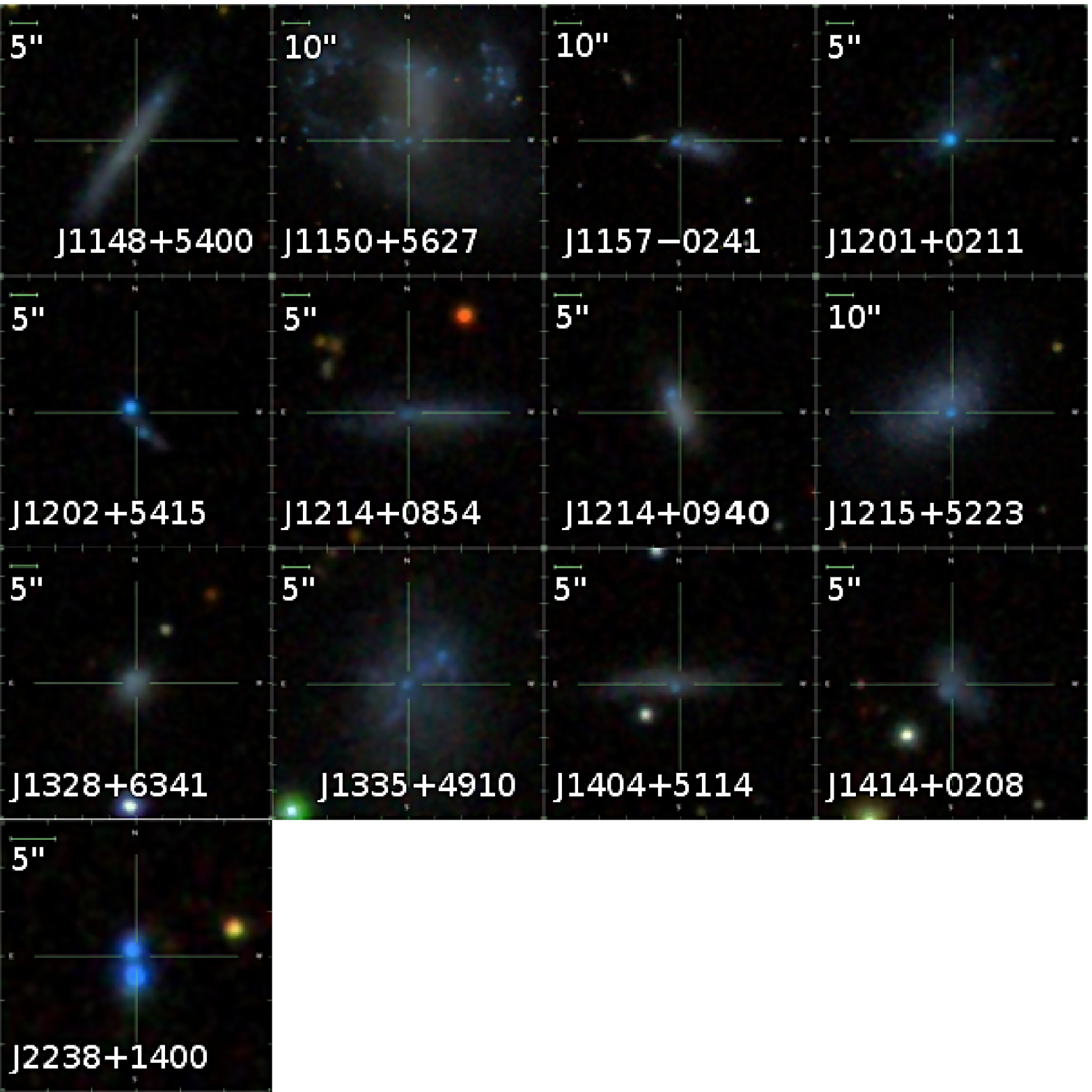}
\caption{(Continued)}
\end{figure*} 

\renewcommand{\thefigure}{\arabic{figure}}

Observed and derived H {\sc i} parameters for all detected 
galaxies (28 objects) are given in Table \ref{tab1}.  
Column 1 lists the galaxies in order of increasing right ascension 
(given in Table \ref{tab2}). Column 2 gives the integrated H {\sc i} flux densities
and their errors. The errors are taken to be equal to \citep{Thu99}

\begin{equation}
\sigma = 2 \times \sqrt{N} \times RMS \times \Delta{v}, 
\end{equation}

\noindent
where $N$ is the number of 
channels over which the line is detected, $RMS$ is the root-mean-square 
deviation in mJy in the baseline fit to the 20-channel 
boxcar smoothed 21 cm spectrum, as given in column 3, 
and $\Delta v$ is the velocity resolution equal to 3.3 km s$^{-1}$.

We can use the measured $RMS$ of 1.86 mJy  
of the spectrum of the non-detected galaxy 
J2238+1400, to derive an upper limit for its H {\sc i} mass.
This galaxy, at the distance of 86.4 Mpc, is the furthest object in our 
sample.  
We take its H~{\sc i} velocity width at zero intensity to be 192 km s$^{-1}$, 
obtained by multiplying 155 km s$^{-1}$, the largest $\Delta v_{20}$ in our 
sample (Table \ref{tab1}), by 1.2, assuming that the velocity width at zero intensity 
is 20\% wider than $\Delta{v_{20}}$. Assuming a rectangular profile, 
we obtain a 5$\sigma$ upper 
limit $M$(H {\sc i}) $\leq$ 3.1 $\times$ 10$^9$ M$_\odot$.



Columns 4 and 5 give respectively the  
velocity widths at 20\% of maximum intensity and the mid-point velocities 
of the 20\% profile widths,  
while columns 6 and 7 list the same quantities at  
50\% of maximum intensity. 
Column 8 lists H~{\sc i} masses calculated from the integrated line flux densities using the equation
\begin{equation}
M {(\rm H I)} = 2.36 \times 10^{5}\int Sdv\,D^{2}\,{\rm M}_{\sun}, 
\end{equation}
where $D$ is the distance in Mpc given in column 3 of Table \ref{tab2} 
and $\int$S\I{dv} is the integrated H {\sc i} 
flux density in Jy km s$^{-1}$, given in column 2.

Column 9 gives the logarithm of the
 ratio of the H {\sc i} mass to absolute \I{g} luminosity of each 
galaxy. The luminosity  is 
given in column 4 of Table \ref{tab2}, and we adopt $M_{g,\odot}$ = 5.12 mag \citep{B05}.


\subsection{Shapes of H {\sc i} profiles and galaxy morphologies}   

Column 10 gives the classification of each H {\sc i} profile. 
We have classified the H {\sc i} profiles into two broad categories, depending on their shapes: the Gaussian (G) profiles characteristic of diskless 
dwarf galaxies 
without large rotational motions, and the steep-sided double-horn (DH) profiles characteristic of inclined disk galaxies with significant  
rotational motions. The reliabilty of the profile classification depends
 on the signal-to-noise 
ratio of the spectrum, decreasing with lower values of that ratio. Most of the H {\sc i} profiles (20 out of 28, or 71\%) have a Gaussian shape,
indicating dominant random motions, while the 
remaining have double-horned profiles, reflecting dominant 
rotational motions.

\setcounter{table}{0}
\begin{table*}
 \centering
  \caption{GBT Objects, HI Data \label{tab1}}
  \begin{tabular}{@{}cccccccccc@{}}
  \hline
   Object & $S\textit dv$ & $RMS$ 
  & {$\Delta v_{20}$} & {\textit{v}$_{20}$} & {$\Delta v_{50}$} & {\textit{v}$_{50}$} 
 & {log $M$(H {\sc i})} & {log $M$(H {\sc i})/$L_g$} &
Profile
  \\
 &{(Jy km s$^{-1}$)} & (mJy)  &{(km s$^{-1}$)} &{(km s$^{-1}$)} &{(km s$^{-1}$)} &{(km s$^{-1}$)} 
 &{(M$_{\odot}$)} &{(M$_{\odot}$ / L$_{\odot}$)}  & Shape
\\
(1) &(2) &(3) &(4) &(5) &(6) &(7) &(8) &(9) &(10) 
\\

\hline 
J0031$-$0934 & 0.790 $\pm$ 0.097 & 2.05 & 134 & 3398 & 75 & 3386  & 8.60 
& 0.400  & DH \\
J0133+1342 & 0.208 $\pm$ 0.078 & 2.95 & 43 & 2596 & 31 & 2598  & 7.80  
& -0.191 &  G \\
J0215$-$0846 & 0.666 $\pm$ 0.056 & 1.85 & 51 & 1480 & 35 & 1481  & 7.78    
& 0.204 &  G \\
J0843+4025 & 0.220 $\pm$ 0.060 & 2.14 & 23 & 629 & 16 & 628  & 6.80  
& -0.256 &  G \\
J0859+3923 & 0.806 $\pm$ 0.088 & 2.68 & 52 & 572 & 39 & 567  & 7.30  
& 0.057 &  G \\
J0903+0548 & 0.297 $\pm$ 0.060 & 1.79 & 75 & 3862 & 51 & 3868  & 8.30 
& 0.013 &  G \\
J0908+0517 & 2.712 $\pm$ 0.109 & 3.38 & 43 & 599 & 27 & 598  & 7.42
& 1.01 &  G \\
J0921+3944 & 2.351 $\pm$ 0.098 & 1.85 & 155 & 4165 & 136 & 4166  & 9.30 
& 0.769 &  DH\\
J0944+0936 & 4.414 $\pm$ 0.128 & 3.07 & 91 & 542 & 70 & 543  & 7.48 
& -0.316 &  G \\
J1044+0353 & 0.373 $\pm$ 0.058 & 1.61 & 51 & 3849 & 41 & 3845  & 8.40 
& -0.099 &  G \\
J1055+5111 & 0.418 $\pm$ 0.060 & 1.55 & 82 & 1394 & 46 & 1378  & 7.74  
& -0.184 &  G \\
J1119+0935 & 1.959 $\pm$ 0.116 & 3.33 & 73 & 990 & 51 & 991  & 7.83 
& 0.187 &  G \\
J1121+0324 & 2.378 $\pm$ 0.118 & 2.75 & 106 & 1175 & 89 & 1176  & 8.35 
& 0.770 &  DH \\
J1121+5720 & 0.424 $\pm$ 0.082 & 2.48 & 67 & 1086 & 56 & 1081  & 7.66 
& -0.107 &  G \\
J1127+6536 & 0.157 $\pm$ 0.071 & 2.00 & 53 & 1246 & 52 & 1247  & 7.25  
& -0.365 &  G \\
J1128+5714 & 0.648 $\pm$ 0.134 & 3.82 & 66 & 1675 & 61 & 1676  & 8.08 
& 0.316 &  DH \\
J1148+5400 & 0.869 $\pm$ 0.091 & 2.01 & 108 & 2614 & 92 & 2619  & 8.51  
& 0.389 & DH \\
J1150+5627 & 12.99 $\pm$ 0.122 & 2.89 & 134 & 888 & 118 & 886  & 8.95 
& 0.731 & DH \\
J1157$-$0241 & 1.065 $\pm$ 0.114 & 3.25 & 44 & 1392 & 32 & 1392  & 8.15 
& 0.548 &  G \\
J1201+0211 & 0.869 $\pm$ 0.104 & 3.42 & 32 & 976 & 26 & 977  & 7.18  
& 0.206 &  G \\
J1202+5415 & 0.093 $\pm$ 0.043 & 1.80 & 32 & 3589 & 29 & 3590  & 7.79   
& -0.231 &  G \\
J1214+0854 & 1.919 $\pm$ 0.120 & 2.98 & 97 & 1928 & 87 & 1929  & 7.89 
& 0.750 &  DH \\
J1214+0940 & 0.118 $\pm$ 0.051 & 2.23 & 28 & 1702 & 26 & 1702  & 6.68 
& -0.688 &  G \\
J1215+5223 & 7.077 $\pm$ 0.089 & 2.56 & 47 & 161 & 30 & 160  & 7.27 
& 0.187 &  G \\
J1328+6341 & 0.193 $\pm$ 0.073 & 1.86 & 59 & 1789 & 13 & 1790  & 7.61 
& -0.152 &  G \\
J1335+4910 & 1.680 $\pm$ 0.094 & 3.03 & 44 & 632 & 27 & 631  & 7.83 
& 0.211 &  G \\
J1404+5114 & 1.008 $\pm$ 0.069 & 1.68 & 103 & 1774 & 92 & 1778  & 8.30 
& 0.415 &  DH \\
J1414$-$0208 & 0.670 $\pm$ 0.075 & 2.29 & 65 & 1556 & 48 & 1554  & 7.99 
& 0.347 & G \\
J2238+1400 & ... & 1.86 & ... & ... & ... & ... & $\leq$9.5 & ... & .. \\  

\hline
\end{tabular}
\noindent Notes: The columns are as follows. (1): Source name. Some of the objects are known under other names: J0944+0936 = IC559; J1202+5415 = SBS1159+545; J1215+5223 = CGCG269-049;  (2): H {\sc i} integrated flux density and error. (3): Root-mean-square deviation in the baseline fit. (4): H {\sc i} line width at 20\% of the peak flux density. (5): Velocity at 20\% of maximum. (6): H {\sc i} line width at 50\% of the peak flux density. (7): Velocity at 50\% of maximum. (8): H {\sc i} gas mass. (9): Ratio of H {\sc i} gas mass to \I{g}-band luminosity. (10):  G= Gaussian profile; DH= Double-horned profile.
\end{table*}

\setcounter{table}{1} 
\begin{table*}
 \centering
  \caption{GBT Objects, Optical Data.} 
  	\label{tab2}
  \begin{tabular}{@{}cccccccccc@{}}
  \hline
  Object & RA  & {$D$(Mpc)} & {\textit{m}$_g$} & {\textit{r}$_g$(\arcsec)} & [O/H] & log SFR & {log $M_{\R{*}}$} & log $M_{\R{dyn}}$(M$_{\odot}$) & {log $\tau$(yr)} 
  \\
  & Dec & {$z$} & {$M_g$} & {$R_g$(kpc)} & {\textit{$i$}($^{\circ}$)} & (M$_{\odot}$ yr$^{-1}$) & {log $M_{\R{y}}$} & {$M$(H {\sc i})/{$M_{\R{dyn}}$}} & \textit{f}$_{\R{gas}}$
\\
 & (J2000) & & & & & log \textit{f}$_{\R{cor}}$ & {(M$_{\odot}$)} & & 
  \\
(1) &(2) &(3) &(4) &(5) &(6) &(7) &(8) &(9) &(10)   
\\ 

\hline
& ~~~\,00:31:40 \vspace{-0.2cm} & 46.2 & 17.94  & 8.07 & 7.55 & -1.05 & 7.719 & 9.75 & 9.565\\
J0031$-$0934 & & & & & & & &  \\
& $-$09:34:34 & 0.0113 & -15.38 & 1.81 & 70.1 & 0.952 & 5.952 & 0.071 & 0.914 \\
\\
& ~~~\,01:33:52 \vspace{-0.2cm} & 36.1 & 17.92 & 5.30 & 7.55 & -1.07 & 6.639 & 8.52 & 8.87 \\
J0133+1342 & & & & & & & &  \\
& +13:42:09 & 0.0087 & -14.87 & 0.93 & 81.9 & 0.488 & 5.743 & 0.190 & 0.954\\
\\
& ~~~\,02:15:13 \vspace{-0.2cm} & 19.5 & 17.64 & 12.10 & 7.58 & -1.93 & 6.826 & 8.72 & 9.71\\
J0215$-$0846 & & & & & & & &  \\
& $-$08:46:24 & 0.0050 & -13.81 & 1.14 & 42.8  & 0.900 & 5.188 & 0.116 & 0.936\\
\\
& ~~~\,08:43:37 \vspace{-0.2cm} & 10.17 & 17.51  & 8.45 & 7.44 & -2.80 & 6.395 & 7.60 & 9.60\\
J0843+4025 & & & &  & & & &  \\
& +40:25:46 & 0.0020 & -12.53  & 0.42 & 58.2  & 0.928 & 4.748 & 0.158 & 0.782\\
\\
& ~~~\,08:59:46 \vspace{-0.2cm} & 10.2 & 17.06  & 10.29 & 7.38 & -2.74 & 7.154 & 8.46 & 10.04\\
J0859+3923 & & & & & & & &  \\
& +39:23:06 & 0.0020 & -12.98  & 0.51 & 42.8  & 1.252 & 4.766 & 0.069 & 0.660\\
\\
& ~~~\,09:03:00 \vspace{-0.2cm} & 53.1 & 18.04 & 8.03 & 7.55 & -1.08 & 7.684 & 9.30 & 9.38\\
J0903+0548 & & & & & & & &  \\
& +05:48:23 & 0.0129 & -15.59 & 2.07 & 65.5 & 0.772 & 5.896 & 0.100 & 0.852\\
\\
& ~~~\,09:08:36 \vspace{-0.2cm} & 6.45 & 16.46 & 15.0 & 7.39 & -3.158 & 5.658 & 8.10 & 10.58\\
J0908+0517 & & & & & & & &  \\
& +05:17:27 & 0.0020 & -12.59 & 0.47 & 56.1 & 0.672 & 3.886 & 0.209 & 0.988\\
\\
& ~~~\,09:21:19 \vspace{-0.2cm} & 59.7 & 17.68 & 7.65 & 7.50 & -1.03 & 7.494 & 10.33 & 10.33\\
J0921+3944 & & & & & &  & &  \\
& +39:44:59 & 0.0140 & -16.20 & 2.21 & 73.7 & 0.736 & 5.877 & 0.093 & 0.989\\
\\
& ~~~\,09:44:44 \vspace{-0.2cm} & 5.39 & 14.29 & 22.0 & 7.49 & ... & ... & 9.02 & ... \\
J0944+0936 & & & & & & & &  \\
& +09:36:49 & 0.0018 & -14.37 & 0.58 & 65.5 & ... & ... & 0.029 & ...\\
\\
& ~~~\,10:44:57 \vspace{-0.2cm} & 53.7 & 17.16 & 5.5 & 7.46 & -0.360 & 6.398 & 8.95 & 8.76\\
J1044+0353 & &  & & & &  & &  \\
& +03:53:13 & 0.0129 & -16.49 & 1.43 & 34.8  & 0.140 & 6.385 & 0.282 & 0.993\\
\\
& ~~~\,10:55:08 \vspace{-0.2cm} & 23.5 & 17.53 & 8.76 & 7.59 & -1.73 & 6.708 & 8.90 & 9.47\\
J1055+5111 & &  & & & & & &  \\
& +51:11:19 & 0.0046& -14.33 & 1.00 & 61.6 & 0.788 & 6.343 & 0.069 & 0.937\\
\\
& ~~~\,11:19:28 \vspace{-0.2cm} & 12.1 & 16.42 & 15.52 & 7.52 & -1.95 & 6.323 & 8.95 & 9.78\\
J1119+0935 & & & & & & & &  \\
& +09:35:44 & 0.0033 & -13.99 & 0.91 & 66.9  & 0.924 & 4.977 & 0.076 & 0.978\\
\\
& ~~~\,11:21:52 \vspace{-0.2cm} & 20.0 & 16.56 & 24.5 & 7.56 & -2.162 & 6.169 & 10.09 & 10.51\\
J1121+0324 & & & & & & & &  \\
& +03:24:21 & 0.0041 & -14.95 & 2.38 & 59.3 & 0.448 & 5.072 & 0.018 & 0.995\\
\\
& ~~~\,11:21:47 \vspace{-0.2cm} & 21.3 & 17.35 & 11.18 & 7.55 & -2.32 & 6.909 & 9.13 & 9.98\\
J1121+5720 & & & & & & & &  \\
& +57:20:48 & 0.0036 & -14.29  & 1.15 & 75.9 & 1.024 & 4.950 & 0.034 & 0.887\\
\\
& ~~~\,11:27:17 \vspace{-0.2cm} & 21.9 & 17.79 & 7.82 & 7.46 & -1.74 & 6.862 & 8.92 & 8.99\\
J1127+6536 & & & & & & & &  \\
& +65:36:03 & 0.0041 & -13.91 & 0.83 & 48.6 & 0.912 & 5.341 & 0.021 & 0.773\\
\\
& ~~~\,11:28:24 \vspace{-0.2cm} & 27.9 & 17.95 & 9.19 & 7.55 & -1.71 & 6.640 & 9.45 & 9.79\\
J1128+5714 & & & & & & & &  \\
& +57:14:48 & 0.0056 & -14.28 & 1.24 & 62.9 & 0.800 & 5.824 & 0.043 & 0.974 \\
\\ \hline
\end{tabular}
\end{table*}

\setcounter{table}{1}
\begin{table*}
 \centering
  \caption{Continued} 
  \begin{tabular}{@{}cccccccccc@{}}
    \hline
   & RA \vspace{-0.3cm} & {$D$(Mpc)} & {\textit{m}$_g$} & {\textit{r}$_g$(\arcsec)} & [O/H] & log SFR & {log $M_{\R{*}}$} & log $M_{\R{dyn}}$(M$_{\odot}$) & {log $\tau$(yr)} 
 \\ \vspace{-0.3cm}
  \\
  Object & & & & & & & & & 
  \\
& Dec & {$z$} & {$M_g$} & {$R_g$(kpc)} & {\textit{$i$}($^{\circ}$)} & (M$_{\odot}$ yr$^{-1}$) & 
{log $M_{\R{y}}$} & {$M$(H {\sc i})/{$M_{\R{dyn}}$}} & \textit{f}$_{\R{gas}}$
 \\
 & (J2000) & & & & & log \textit{f}$_{\R{cor}}$ & {(M$_{\odot}$)} & & 
  \\
(1) &(2) &(3) &(4) &(5) &(6) &(7) &(8) &(9) &(10)   
\\ 

\hline
& ~~~\,11:48:01 \vspace{-0.2cm} & 39.7 & 17.13 & 13.7 & 7.54 & ... & ... & 10.11  & ... \\
J1148+5400 & & & & & & & & \\
& +54:00:19 & 0.0086 & -15.86 & 2.64 & 65.5 & ... & ... & 0.025 & ...\\
\\
& ~~~\,11:50:47 \vspace{-0.2cm} & 17.0 & 15.73 & 21.78 & 7.47 & -1.59 & 7.415 & 10.80 & 10.54\\
J1150+5627 & & & & & & & &  \\
& +56:27:06 & 0.0030 & -15.42 & 1.80 & 26.1 & 1.660 & 5.681 & 0.014 & 0.909\\
\\
& ~~~\,11:57:12 \vspace{-0.2cm} & 23.8 & 17.99 & 13.29 & 7.57 & -1.69 & 6.379 & 8.77 & 9.84\\
J1157$-$0241 & & & & & & & &  \\
& $-$02:41:11 & 0.0047 & -13.89 & 1.53 & 56.1 & 0.636 & 5.868 & 0.240 & 0.988 \\
\\
& ~~~\,12:01:22 \vspace{-0.2cm} & 8.6 & 17.70 & 11.71 & 7.51 & -1.20 & 6.131 & 8.09 & 8.38\\
J1201+0211 & & & & & & & &  \\
& +02:11:08 & 0.0033 & -11.97 & 0.49 & 51.3 & 0.688 & 5.643 & 0.123 & 0.940 \\
\\
& ~~~\,12:02:02 \vspace{-0.2cm} & 53.3 & 18.21& 5.8 & 7.50 & -0.788 & 6.295 & 8.67 & 8.58\\
J1202+5415 & & & & & & & &  \\
& +54:15:50 & 0.0120 & -15.42 & 1.50 & 28.7 & 0.192 & 5.892 & 0.132 & 0.978 \\
\\
& ~~~\,12:14:13 \vspace{-0.2cm} & 13.1 & 17.86 & 10.84 & 7.57 & -1.25 & 7.053 & 9.41 & 9.14\\
J1214+0854 & & & & & & & &  \\
& +08:54:30 & 0.0064 & -12.73 & 0.69 & 81.9 & 1.204 & 5.821 & 0.030 & 0.906\\
\\
& ~~~\,12:14:53 \vspace{-0.2cm} & 13.1 & 17.29 & 5.87 & 7.55 & -2.05 & 7.258 & 7.97 & 8.73\\
J1214+0940 & & & & & & & &  \\
& +09:40:11 & 0.0056 & -13.30 & 0.37 & 61.6 & 0.848 & 5.813 & 0.051 & 0.270\\
\\
& ~~~\,12:15:46 \vspace{-0.2cm} & 3.33 & 15.03 & 30.62 & 7.47 & -3.49 & 6.027 & 8.22 & 10.76\\
J1215+5223 & & & & & & & &  \\
& +52:23:14 & 0.005 & -12.58 & 0.49 & 12.1 & 1.532 & 5.055 & 0.112 & 0.961 \\
\\
& ~~~\,13:28:22 \vspace{-0.2cm} & 29.8 & 18.09 & 6.87 & 7.56 & -1.71 & 7.033 & 7.80 & 9.32\\
J1328+6341 & & & & & & & &  \\
& +63:41:07 & 0.0060 & -14.28 & 0.99 & 39.8 & 0.724 & 6.378 & 0.631 & 0.840\\
\\
& ~~~\,13:35:42 \vspace{-0.2cm} & 13.01 & 16.23 & 10.38 & 7.60 & -3.09 & 5.177 & 8.25 & 10.92\\
J1335+4910 & & & & & & & &  \\
& +49:10:35 & 0.0022 & -14.34 & 0.65 & 30.7 & 0.168 & 3.838 & 0.380 & 0.998\\
\\
& ~~~\,14:04:32 \vspace{-0.2cm} & 29.1 & 17.72 & 11.87 & 7.58 & -1.70 & 7.096 & 9.84 & 10.00\\
J1404+5114 & & & & & & & &  \\
& +51:14:05 & 0.0058 & -14.60 & 1.67 & 81.9 & 1.004 & 5.263 & 0.029 & 0.958\\
\\
& ~~~\,14:14:54 \vspace{-0.2cm} & 24.9 & 17.99 & 7.33 & 7.35 & -1.86 & 7.056 & 8.88 & 9.85\\
J1414$-$0208 & & & & & & & &   \\
& $-$02:08:22 & 0.0051 & -13.99 & 0.88 & 41.3 & 0.932 & 5.447 & 0.129 & 0.923\\
\\ \hline
\end{tabular}
\\
\noindent Notes: The columns are as follows. (1): Source name. (2): Right ascension and declination. (3): Distance in megaparsecs and optical redshift. (4): Apparent and absolute SDSS \I{g}-band Petrosian magnitude. (5): Optical major-axis radius, in arcseconds and kiloparsecs, containing 90\% of the g-band light. From the SDSS. (6) Oxygen abundance 12+log O/H and inclination angle of the source derived from the SDSS axial ratio. (7): Logarithm of the star formation rate. (8): Logarithm of the young and total stellar masses. (9): Logarithm of the dynamical mass derived from the \I{g}-band Petrosian radius. (10): Logarithm of the depletion time scale and the gas mass fraction. 
\end{table*}

We have examined the morphology 
of all galaxies in the sample using the SDSS images (Fig. \ref{fig2}). Using the classification terminology of \citet{San16},
we find 3 galaxies (J0133+1343, J1055+5111 and J1201+0211) or 10\% of the sample to be symmetric, 
with a centrally concentrated emission, 3 galaxies (J1044+0353, J1202+5415 and J2238+1400) possessing 2 knots (10\%), 5 galaxies (J0908+0517, J0944+0936, J1148+5400, J1150+5627 and J1335+4910) having multiple knots (17\%) and the remaining 18 galaxies (63\%) having a cometary structure.
According to the 
BCD morphological classification scheme of \citet{LT85}, the cometary BCDs 
are characterized by a high surface brightness star-forming region 
(the comet's ``head'') located near one end of an elongated lower surface brightness stellar body (the comet's ``tail''). The latter could be interpreted as 
a rotating thin stellar disk 
seen nearly edge-on, but with modest 
rotational velocities and significant random motions, as suggested by 
the predominance of the H {\sc i} profiles with a Gaussian shape.  
It is interesting that the low-metallicity selection criterion 
has picked out a majority of cometary BCDs. How does the  
optical morphology classification of the GBT sample compare with  
that of other XMD samples such as the one assembled 
by \citet{Mor11}? These authors also  
find in their XMD sample 
a predominance of cometary morphology (75\%), the remaining 25\% showing a 
single knot without any obvious underlying low-surface brightness component.
\citet{San16} found in their XMD sample that 57\% of the galaxies are cometary, with 23\% symmetric,
10\% multi-knot and 4\% 2-knot. The common feature of all these XMD samples is that they all show a large proportion (more than half) of cometary galaxies.   
Recently, 
\citet{San15} have suggested that these metal-deficient cometary galaxies 
are the result of  
low-metallicity gas clouds falling onto low-surface-brightness galaxy 
disks, 
and triggering  
bursts of star formation. The star-forming 
regions tend to be at the ends of the low-surface-brightness disks 
because the impacting gas clouds have the largest effects on 
the outer parts of galaxies where the ambient pressure and column density 
are low.


\begin{figure}
\centering
\includegraphics[scale=0.35]{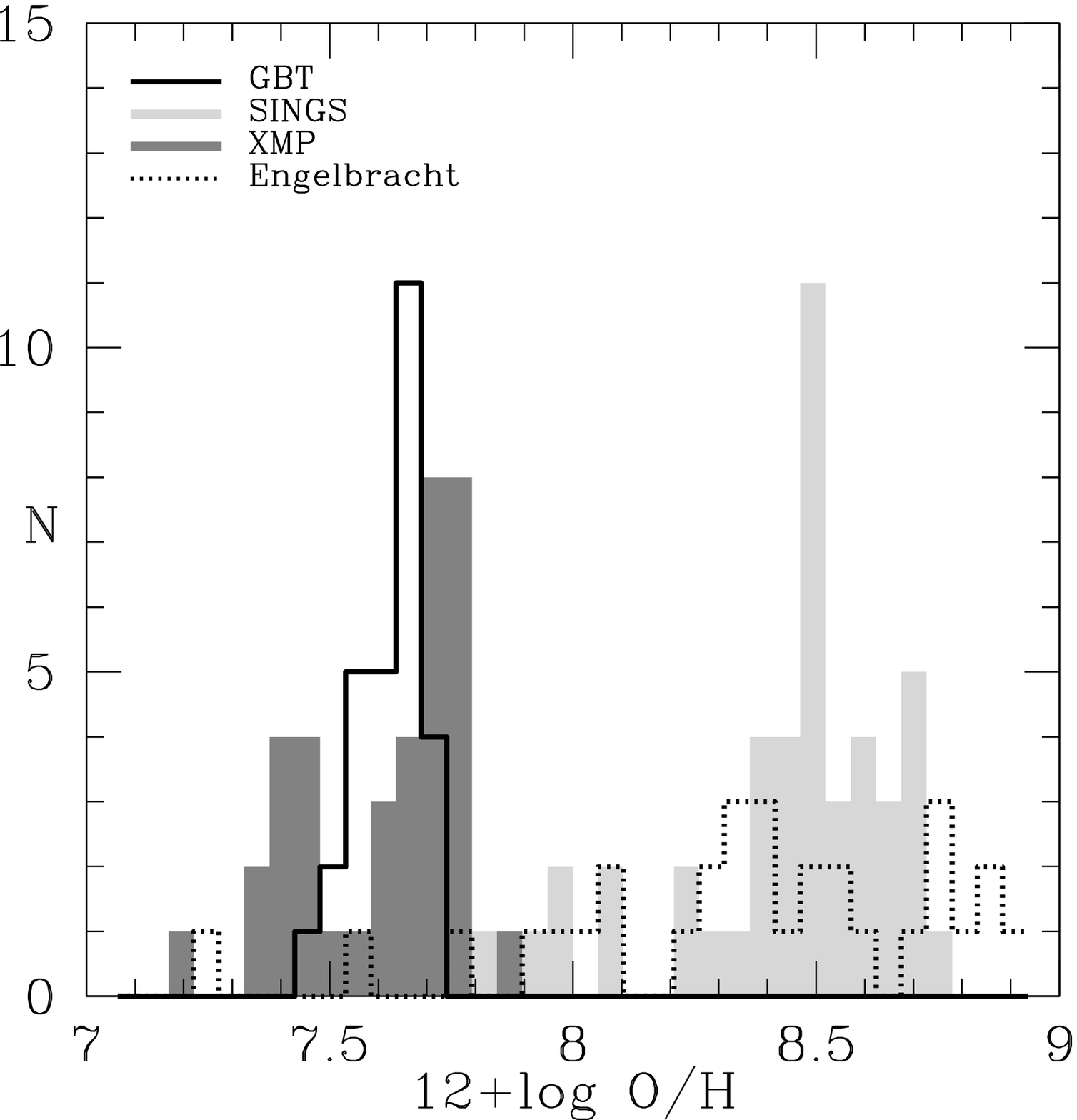}
\caption{Histogram of oxygen abundance for the GBT and the 3 comparison 
XMP, SINGS and Engelbracht samples.} 
\label{fig3}
\end{figure}

\subsection{Comparison with previous H {\sc i} observations}
A few galaxies in our sample have had previous H {\sc i} observations. 
They are J0133+1342, J1121+0324, J1201+0211, 
J1215+5233 and J1202+5415 \citep{Fil13}, J0944+0936 \citep{Sti09}, 
J0908+0517 and J1119+0935 \citep{Pop11}, J1121+0324, J1201+0211 and 
J1215+5223 \citep{Pus07}, and J1214+0940 \citep{Kent08}. 
Within the errors, our flux densities
 are in good general agreement with those of 
previous authors, except for two objects: J1215+5223 for which 
\citet{Fil13} found 4.7$\pm$0.5 
Jy km s$^{-1}$ and \citet{Pus07} found 5.2$\pm$0.2 Jy km s$^{-1}$
as compared to our flux density 
of 7.1 Jy km s$^{-1}$, and J1119+0935 for which      
 \citet{Pop11} found a flux density of 1.4$\pm$0.1 Jy km s$^{-1}$ 
as compared to our value of 1.9$\pm$0.1 Jy km s$^{-1}$. These measurements 
were made with telescopes having similar beam sizes so the cause of the discrepancies is not clear.

\section{Supplemental data}

We have also listed in Table \ref{tab2} 
other data, useful for characterizing the evolutionary 
state and star-forming properties of the 28 detected galaxies.  
Column 2 gives the J2000 coordinates of each galaxy. 
Column 3 gives the redshift as obtained 
from the emission lines in the 
optical spectra (lower line) and the distance derived from the redshift 
corrected for Virgo infall (upper line)
 Column 4 gives the apparent (upper line) and absolute 
(lower line) $g$-band magnitudes 
within the galaxy's Petrosian 
90\% radius, as taken from the SDSS DR10. Column 5 lists the Petrosian 
90\% angular (upper line) and linear (lower line) radii,    
as taken from the SDSS DR10. 
To check the reliability of the photometric measurements of magnitudes and radii of the 
SDSS DR10, we have inspected visually all objects in the GBT sample and derived independently their 
$g$ angular radii $r_g$. For 21 out of 28 objects (75\%), our measured radii agree to within 10\% with the SDSS DR10 radii. However, there are large discrepancies for 7 objects, the SDSS radius being too small as compared to our measured radius. These objects are J0908+0157, J0944+0936 = IC 559, J1044+0353, J1121+0324, J1148+5400, J1202+5415 and J1335+4910. With the exception of J1121+0324 that has a cometary structure, all of these objects have either a 2- or multi-knot structure.
Evidently, the SDSS surface photometry routine does not handle well a multi-knot structure and derives radii 
and magnitudes for a single H {\sc ii} region rather than for the whole object. 
For these 7 objects, 
to make sure that our derived $r_g$ refers not only to the size of an individual H {\sc ii}  region, but to that of the whole galaxy, we have derived by surface photometry our own $g$ angular radii containing 90\% of the total light and the magnitudes within these radii using the IRAF/APPHOT routine~\footnote{IRAF is distributed by the National Optical Astronomy Obser-
vatories, which are operated by the Association of Universities for
Research in Astronomy, Inc., under cooperative agreement with
the National Science Foundation}.  


Column 6 (upper line) gives for each galaxy 
the oxygen abundance [O/H]. As our galaxies were 
selected to possess a well-detected [O~{\sc iii}] $\lambda$4363 emission 
line in their spectra, these abundances are well determined, using the direct 
method. The derived oxygen abundances are generally those of the brightest H~{\sc ii} region in the galaxy, typically the SDSS spectroscopic target. This however is a good indicator of the metallicity of the whole BCD as detailed spatial metallicity studies of a few of these objects show no evidence of large metallicity 
gradients \citep*[e. g. ][]{Thu99a,Noe00}.

Column 6 (lower line) gives the 
the inclination angle $i$ in degrees of the plane of the 
galaxy to the plane of the sky ($i$ = 0 for face-on and $i$ = 90$^\circ$
 for edge-on).
Following \citet{Thu81},
$i$ is calculated from

\begin{equation}
\cos^{2} i =\frac{r_{25}^{2} - r_{0}^{2}}{1-r_{0}^{2}},  
\end{equation}

\noindent
where $r_{25}$ is the axial ratio as obtained from the SDSS DR10 and  
$r_0$ is the intrinsic axial ratio of the galaxy. 
Following \citet{Fil13},
we adopt $r_0$=0.25,
i.e. if a galaxy's disk appears more than four times 
elongated than wide, then its inclination angle is set to 90$^\circ$. 
This relatively large value is motivated by the work of \citet*{San10}
who identified a limiting stellar mass $M_*$ $\sim$ 2 $\times$ 10$^9$ 
M$_\odot$ below which low-mass galaxies become systematically thicker because 
of the increasing importance of turbulent motions 
relative to rotational motions.
The stellar masses of the GBT galaxies are all lower than this limiting mass,
so this value is appropriate.  
  
 Column 7 gives the galaxy's present star formation rate (SFR) 
as derived from the extinction-corrected H$\alpha$ fluxes using 
\citet{Ken98}'s relation:

\begin{equation}
{\rm SFR} (\R{M}_{\odot} \; \R{yr}^{-1})=\frac{L({\rm H}\alpha)}{1.26 \times 10^{41} \; \R{erg \; s}^{-1}}\, \,  .
\end{equation}

\noindent
As the H$\alpha$ fluxes have been derived from SDSS spectra obtained 
with fibers of 3\arcsec diameter, we need to correct these SFRs for aperture 
effects. We have multiplied each SFR by a correction factor $f$ equal 
to the ratio of the total flux to the flux within a 3\arcsec\ aperture:

\begin{equation}
f_{cor} = 10^{-0.4 \times [m_r(tot) - m_r(3\arcsec)]} \label{eq5},
\end{equation}
\noindent
where $m_r(tot)$ and $m_r(3\arcsec)$ are respectively the 
$r$ magnitudes within the Petrosian 90\% radius and 3\arcsec\ fiber diameter, 
as given in the SDSS (we have used $r$ magnitudes because the Finding Chart 
site of the SDSS DR10 gives 
magnitudes within the 3\arcsec\ fiber diameter only for the $r$ band). The logarithm of $f_{cor}$ is given in the lower line of column 7. Except for four galaxies, all have log~$f_{cor}$ $\leq$ 1.0. The two galaxies J0944+0936 and J1148+5400 have a large difference between $m_r(tot)$ and $m_r(3\arcsec)$ (more than 4.5 mag), and hence large uncertain aperture corrections. We have thus decided not to include their SFRs in our statistical studies.

Column 8 gives the total stellar mass $M_*$ 
and the young 
(age$<$ 10 Myr) stellar mass $M_{\rm y}$ 
in units of solar masses, derived by model-fitting 
the spectral energy distribution (SED) of each galaxy, as described 
in \citet*{Izo11} and more recently in \citet{Izo16}. 
As all our objects show strong 
line emission, care was taken in deriving these stellar masses to subtract 
the ionized gas emission. Neglecting this correction 
would result in a significant overestimate of the galaxy stellar mass. 
As the $M_*$ are derived from spectra obtained with 3\arcsec\ diameter fibers,
we have also applied to each  $M_*$ and $M_{\rm y}$ the upwards correction factor defined 
by Eq. \ref{eq5}. Like the SFRs, the $M_*$ of the galaxies J0944+0936 and J1148+5400 are not included in our statistical studies because of their large uncertain aperture corrections.

To check the reliability of our derived SFRs and $M_*$ concerning the statistical studies in this paper, we have compared them with those obtained by the Max Planck Institute for Astrophysics -- Johns Hopkins University (MPA-JHU) group (\citet{Kau03} and \citet{Bri04}). This constitutes a good check as the MPA-JHU group derives SFRs and $M_*$ from total photometric magnitudes and thus does not have to apply aperture corrections as done here. We have found a very good 
statistical relation between the MPA-JHU SFRs and $M_*$ and our aperture-corrected quantities. In a plot ot $M_*$ (MPA-JHU) vs. $M_*$(our group), the points scatter nicely along the equality line, with a dispersion of $\sim$0.6 in log $M_*$. The same holds for SFRs, with a dispersion of 0.2 in log SFR. We thus conclude that our aperture-corrected quantities are quite adequate for the statistical studies described here and do not introduce any systematic bias. 
 

Column 9 (upper line) gives an estimate of the dynamical mass $M_{dyn}$ 
of each galaxy. 
We emphasize that these dynamical masses are only approximate. 
However they may help to reveal potentially interesting trends.       
$M_{dyn}$ is 
calculated in the following way.
For galaxies with rotation, characterized by a double-horned H {\sc i} 
 profile, the dynamical mass is estimated according to the following 
equation, derived by assuming an object in dynamical virial equilibrium where 
the gravitational force is balanced by the centrifugal force due to rotation:

\begin{equation}
M_{dyn}({\rm M}_{\odot}) = F \times 2.3 \times 10^{5} v_{c}^{2} (\mathrm{km \, s}^{-1}) \, R({\rm HI})\, (\mathrm{kpc}).
\end{equation}

\noindent
Here $F$ is a multiplicative factor that takes into account 
the degree of flattening of the galaxy. \citet{Leq83} has shown that $F$ varies 
between 0.6 for a flat disk system and 1 for a spherical system. We adopt here 
$F$ = 0.7 which corresponds to a flat disk with a flat rotation curve. 
$R$(H {\sc i}) is the radius of the 
H~{\sc i} gas component. 
It is generally larger than the optical radius of the stellar component 
in the case 
of dwarf galaxies. From interferometric H {\sc i} maps of several BCDs, \citet{Thu81} 
have derived the approximate relation 

\begin{equation}  
R({\rm HI}) =  3 \times R_{g} ,
\end{equation}

\noindent
which we adopt. Here, $R_g$ is 
the Petrosian \textit{g}-band 90\% linear radius (column 5). 
\noindent
$v_c$ is the galaxy's rotational velocity corrected for 
inclination $i$ (column 6):  

\begin{equation} 
v_c = \frac{v}{\sin i} = \frac{\Delta{v_{50}}/2}{\sin i} ,
\end{equation}

\noindent
where $v$ is the galaxy's observed rotational velocity, 
set to be equal to half of the velocity width at 50\% 
of maximum intensity (column 6 of Table \ref{tab1}).

For galaxies with weak rotation, in virial equilibrium, supported against 
gravitational collapse by random motions,  
and characterized by a Gaussian H {\sc i} profile,   
we estimate the 
dynamical mass by: 

\begin{equation}
M_{dyn}({\rm M}_{\odot}) = 2.3 \times 10^{5} \sigma^{2} (\mathrm{km \, s}^{-1}) \, R({\rm HI}) (\mathrm{kpc}).
\end{equation}
Here $\sigma$ is the gas velocity dispersion where 
\begin{equation}
\sigma  = \sqrt{3} \times \Delta{v_{50}}/2.36.
\end{equation}
The factor $\sqrt{3}$  converts the 1-dimensional 
velocity dispersion based on the H {\sc i} velocity width
to a 3-dimensional velocity dispersion, and the factor 1/2.36 converts the 
Full Width at Half Maximum (FWHM) $\Delta{v}$ 
of the Gaussian profile to a velocity 
dispersion. Column 9 (lower line) gives the ratio of the H {\sc i} mass 
to the dynamical mass.   

Finally, column 10 (upper line) gives the gas depletion timescale $\tau$ = 
$M$ (H {\sc i})/SFR 
in units of log (yr).
Column 10 (lower line) gives the gas mass fraction defined as 

\begin{equation}
f_{gas} = \frac{M_{gas}}{M_{baryon}},
\label{gmf}
\end{equation}
where 
\begin{equation}
M_{gas} = 1.4 \times M({\rm HI}).
\end{equation}

The multiplicative factor 1.4 takes 
into account the masses of helium and metals. We do not include 
a correction for molecular hydrogen (H$_2$) as we have no H$_2$
observational constraints in these galaxies, and they are 
almost certainly H {\sc i}-dominated. 

M$_{baryon}$ is the baryonic mass,
defined as the sum of the gaseous and stellar masses:
\begin{equation}
 M_{baryon} = M_{gas} + M_* .
\end{equation}

\section{Comparison samples}

One of the main aims of this paper is to study how the H~{\sc i} 
content of a galaxy varies statistically with 
various global galaxian properties such as metallicity, 
SFR or stellar mass. To this end, the 28 galaxies in our GBT sample are not sufficient. As mentioned 
before, this sample spans only a very narrow range of oxygen abundances: 7.35 $\leq$ [O/H] $\leq$ 7.60. 
We need to supplement our 
data on the GBT sample with data on  
other galaxy samples that span a larger metallicity range. Because 
of the mass-metallicity relation of galaxies, this means that the galaxian 
mass range will also be extended. We have thus searched the literature for 
other galaxy samples that possess data of the same nature as 
the GBT sample, but which span a larger range in metallicity and stellar mass,
so that statistical trends can be studied. We have 
chosen three additional samples for comparison with the GBT sample. 

The first sample consists of a subsample of the extremely metal-poor 
(XMP) galaxy sample of \citet{Fil13} and \citet{Mor11}. Out of the 
140 objects studied by these authors, there are 39 galaxies with the 
necessary ancillary data (oxygen abundances, derived using 
either the direct or strong-line methods, and H {\sc i} fluxes), forming  
what will be called hereafter the `XMP sample'. This sample  
extends the metallicity range at the low-end (to [O/H] = 7.07) and augments
the number of very low-metallicity galaxies, leading to better statistics 
in the low-metallicity range. The relevant data for the XMP sample 
is given in Table \ref{tab3}.

 \begin{figure}
\includegraphics[scale=0.35]{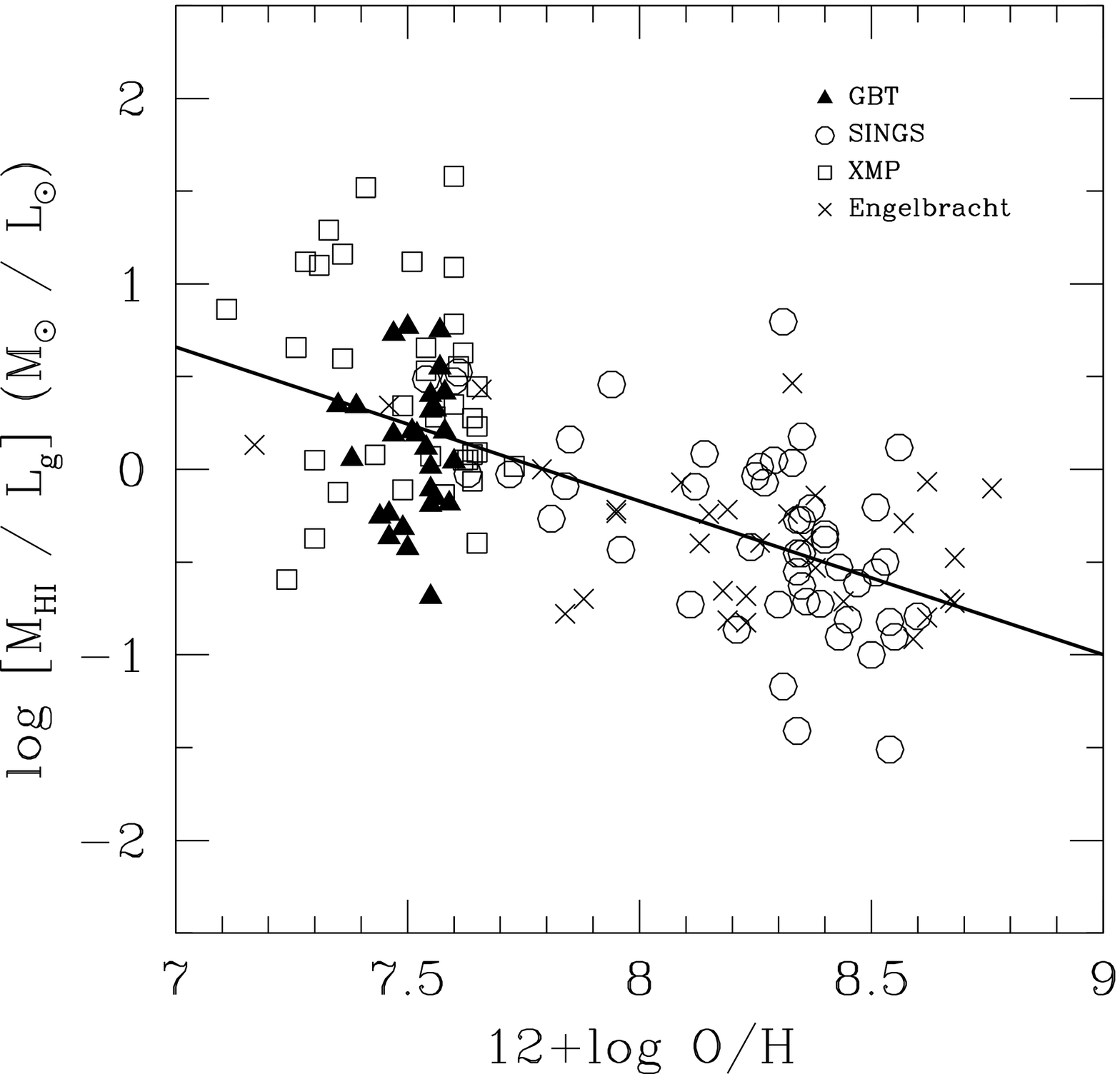}
\caption{Plot of $M$(H {\sc i})/$L_g$ versus oxygen abundance for the GBT and 
the 3 comparison samples. The solid line shows the least-square fit.}
\label{fig4}
\end{figure}

\begin{figure}
\includegraphics[scale=0.35]{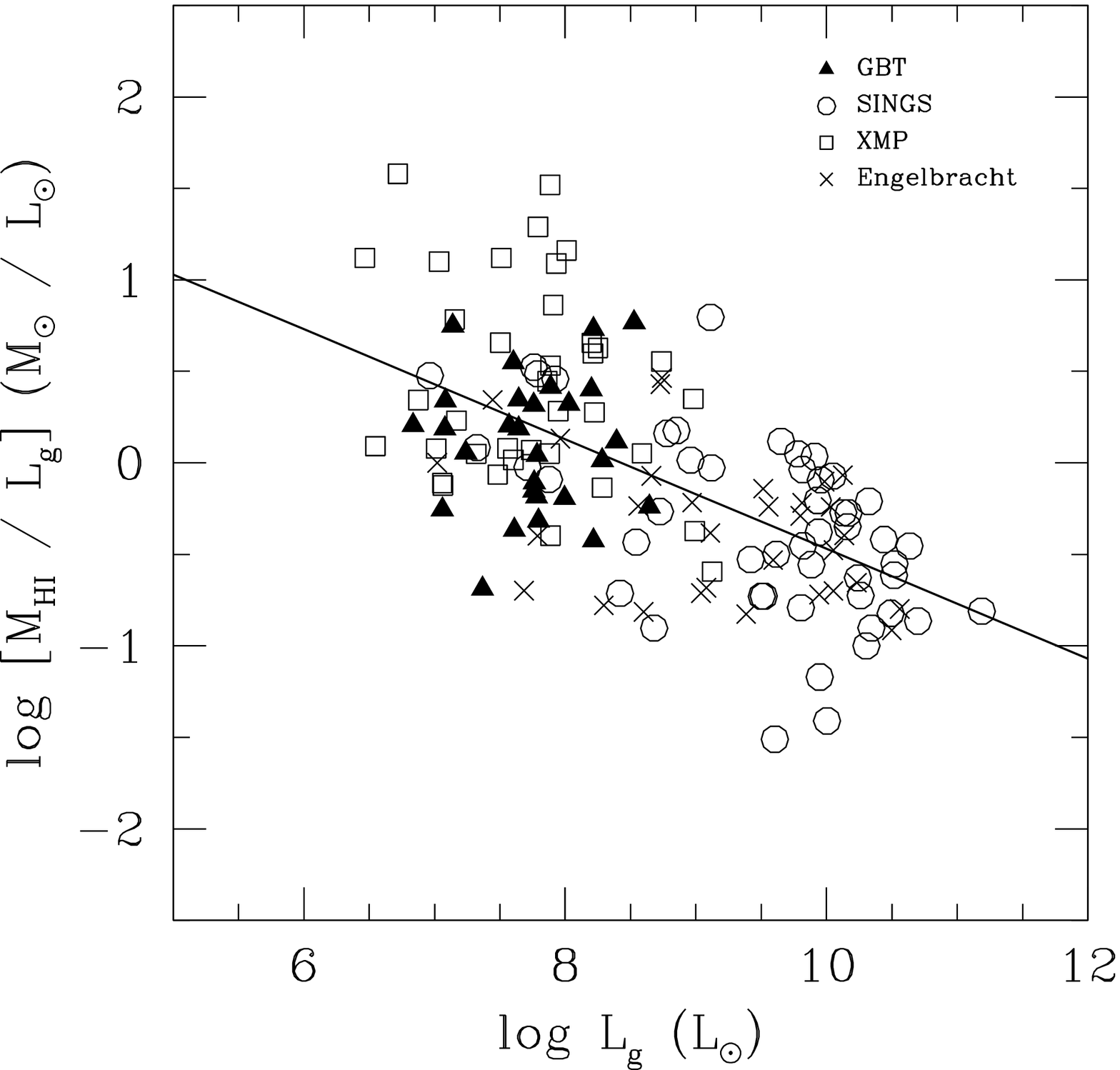}
\caption{Plot of $M$(H {\sc i})/$L_g$ versus absolute $g$ magnitude 
for the GBT and 
the 3 comparison samples. The solid line shows the least-square fit.}
\label{fig5}
\end{figure}

\begin{figure}
\includegraphics[scale=0.35]{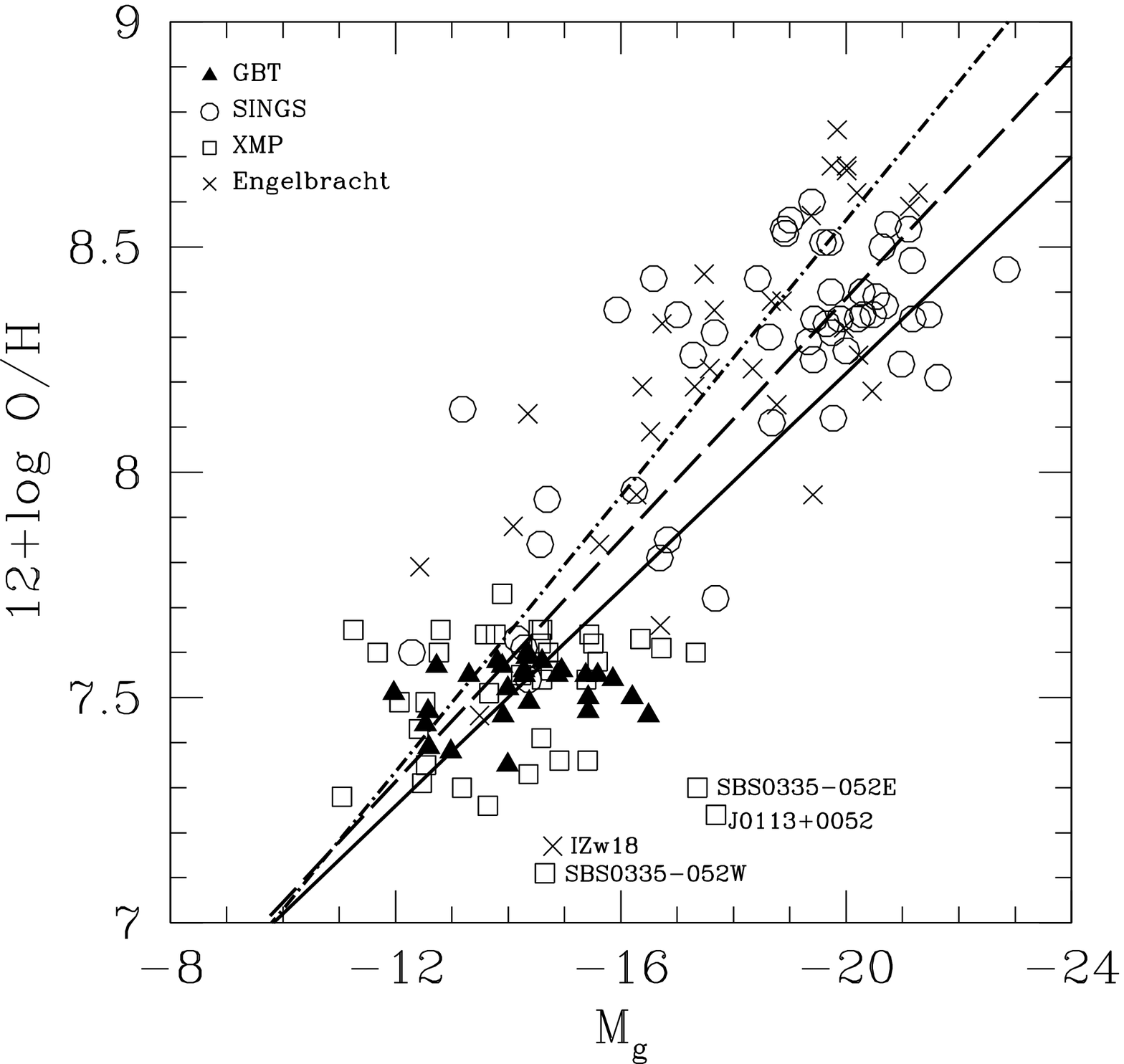}
\caption{Plot of oxygen abundance versus absolute $g$ magnitude for the 
GBT and the 3 comparison samples.
The solid line represents the linear least-squares fit to all 
galaxies. For comparison, the metallicity-luminosity 
relations obtained by \citet{Ski89} (dash-dot line) and 
by \citet{Gus09} (dashed line) are also shown.}
\label{fig6}
\end{figure}

\begin{figure}
\includegraphics[scale=0.4]{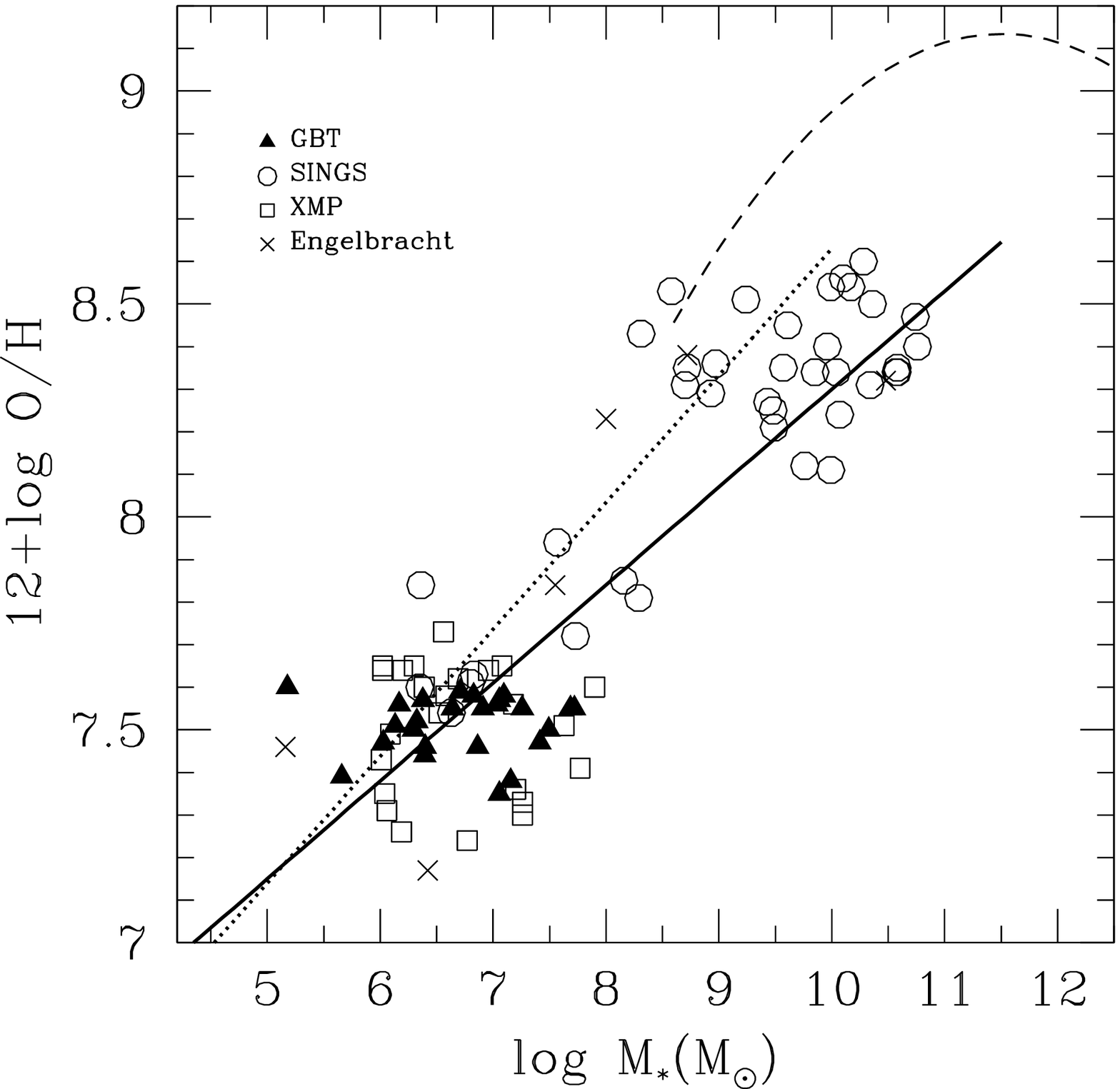}
\caption{Plot of oxygen abundance versus stellar mass. The solid line represents the linear least-square fit to all galaxies in the GBT and the 3 comparison samples. The mass-metallicity relation derived by \citet{Tre04} (dashed line) 
and by \citet{Lee06} (dotted line) are also shown.}
\label{fig7}
\end{figure}

\setcounter{table}{2}
\begin{table*}
 \centering
  \caption{XMP Objects}
  	\label{tab3}
  \begin{tabular}{@{}lcccccccc@{}}
  \hline
   \multicolumn{1}{c}{Object} & $D$(Mpc) &  $M$(H {\sc i}) (M$_{\odot}$) & $m_g$ & $M_g$ & log $M$(H {\sc i})/$L_g$ & [O/H] & log $M_{*}$(M$_{\odot}$) & log SFR(M$_{\odot}$ yr$^{-1}$) 
\\
\multicolumn{1}{c}{(1)} & (2) & (3) & (4) & (5) & (6) & (7) & (8) & (9)
\\  

\hline
J0113+0052 & 15.80 &  3.39E+8 & 13.28$^{a}$ & -17.69 &  -0.594 & 7.24 & 6.77 & ...
\\
J0119$-$0935 & 24.80 &  1.38E+8 & 19.5 & -12.47 & 1.10 & 7.31 & 6.06 & -2.14
\\
HS0122+0743 & 40.30 &  2.14E+9 & 15.7 & -17.33 &  0.350 & 7.60 & 6.39 & -1.17
\\
J0126$-$0038 & 25.80 &  4.27E+8 & 18.4 & -13.66 &  1.12 & 7.51 & 7.63 & -2.93
\\
UGCA20 & 8.63 &  2.00E+8 & 18.0 & -11.68 &  1.58 & 7.60 & ... & ...
\\
UM133 & 22.40 &  4.37E+8 & 15.4 & -16.35 &  0.052 & 7.63 & ... & ...
\\
J0204$-$1009 & 25.20 &  1.48E+9 & 17.1 & -14.91 &  1.16 & 7.36 & 7.20 & -1.54
\\
J0205$-$0949 & 25.30 &  1.95E+9 & 15.3 & -16.72 &  0.554 & 7.61 & ... & ...
\\
J0315$-$0024 & 90.90 &  2.51E+9 & 20.2 & -14.59 &  1.52 & 7.41 & 7.77 & ...
\\
UGC2684 & 5.95 &  8.71E+7 & 16.10$^{b}$ & -12.77 &  0.784 & 7.60 & ... & ...
\\
SBS0335$-$052W & 53.80 &  5.89E+8 & 19.0 & -14.65 &  0.862 & 7.11 & ... & ...
\\
SBS0335$-$052E & 54.00 &  4.17E+8 & 16.3 & -17.36 &  -0.372 & 7.30$^{c}$ & 7.26$^{c}$ & ...
\\
ESO489-G56 & 4.23 &  8.91E+6 & 15.6 & -12.53 &  -0.110 & 7.49 & ... & ...
\\
HS0822+3542 & 11.72 & 8.71E+6 & 17.8 &  -12.54 &  -0.123 & 7.35 & 6.04 & -1.99
\\
HS0846+3522 & 36.30  & 3.09E+7 & 18.2 & -14.60 &  -0.398 & 7.65 & ... & ...
\\
J0940+2935 & 7.23 &  2.51E+7 & 16.5 & -12.80 &  0.231 & 7.65 & 6.30 & -2.09
\\
LeoA & 1.54 &  2.34E+7 & 12.76$^{d}$ & -13.18 &  0.049 & 7.30 & ... & ...
\\
KUG1013+381 & 19.90  & 1.41E+8 & 15.9 & -15.59 &  -0.135 & 7.58 & 6.58 & -1.46
\\
UGCA211 & 15.50  & 1.70E+8 & 16.2 & -14.75 &  0.282 & 7.56 & 7.18 & -2.20
\\
HS1033+4757 & 25.60 &  2.04E+8 & 17.5 & -14.54 &  0.446 & 7.65 & 7.08 & -1.89
\\
HS1059+3934 & 48.10 &  7.59E+8 & 17.9 & -15.51 &  0.628 & 7.62 & ... & ...
\\
J1105+6022 & 23.30 &  3.16E+8 & 16.4 & -15.44 &  0.276 & 7.64 & 6.96 & -1.53
\\
SBS1129+576 & 26.40 &  6.46E+8 & 16.7 & -15.41 &  0.598 & 7.36 & ... & ...
\\
J1201+0211 & 8.60 &  1.66E+7 & 17.6 & -12.07 &  0.344 & 7.49 & 6.09 & -1.99
\\
SBS1121+540 & 17.20 &  4.37E+7 & 17.4 & -13.78 &  0.080 & 7.64 & 6.02 & -1.94
\\
J1215+5223 & 3.33 &  1.23E+7 & 15.2 & -12.41 &  0.078 & 7.43 & 6.01 & -2.82
\\
VCC0428 & 13.10 &  2.63E+7 & 17.0 & -13.59 &  -0.064 & 7.64 & 6.20 & -1.92
\\
Tol65 & 37.90 &  7.24E+8 & 17.5 & -15.39 &  0.656 & 7.54 & ... & ...
\\ 
J1230+1202 & 13.10 &  4.17E+7 & 16.7 & -13.89 &  0.016 & 7.73 & 6.56 & -1.69
\\
UGCA292 & 3.41 &  3.89E+7 & 16.61$^{e}$ & -11.05 &  1.12 & 7.28 & ... & ...
\\
GR8 & 1.43 &  4.37E+6 & 14.53$^{d}$ & -11.25 &  0.092 & 7.65 & 6.02 & -3.16
\\
HS1442+4250 & 12.58 &  2.63E+8 & 15.9 & -14.60 &  0.532 & 7.54 & 6.52 & -2.08
\\
HS1704+4332 & 33.60 &  6.46E+7 & 18.4 & -14.23 &  0.070 & 7.55 & ... & ...
\\
J2053+0039 & 56.40 &  1.20E+9 & 19.4 & -14.36 &  1.29 & 7.33 & 7.26 & -1.73
\\
J2104-0035 & 20.30 &  1.45E+8 & 17.9 & -13.64 &  0.657 & 7.26$^{f}$ & 6.19 & -2.01
\\
J2150+0033 & 63.30 &  1.05E+9 & 19.3 & -14.71 &  1.09 & 7.60 & 7.90 & ...
\\
PHL293B & 22.70 &  8.51E+7 & 17.2 & -14.58 &  0.050 & 7.62 & 6.69 & -1.52
\\
\hline
\end{tabular}
\\
\noindent Notes: The columns are as follows. (1): Source name. Some objects have other names: J0113+0052 = UGC772. (2): Distance. (3): H {\sc i} gas mass. (4): Apparent \I{g}-band magnitude. Magnitudes are from 
\citet{Fil13} unless the object is annotated with a superscript. 
In these cases,
the \textit{g}-mag is derived using a $B-V$ color transformation as described in the text, with the $B$ and $V$ magnitudes from the references listed below.
(5): Absolute \textit{g}-band magnitude. (6): Logarithm of the ratio of H {\sc i} gas mass to \I{g}-band luminosity. (7): Oxygen abundance 12+log O/H. (8): Logarithm of the stellar mass. (9): Logarithm of the star formation rate. 
All data are from \citet{Fil13} unless annotated with a superscript: 
\I{a}) \citet{Pat96}; \I{b}) \citet{Zee96}; \textit{c}) \citet{Izo14}; \I{d}) \citet{deVac}; \I{e}) \citet{Coo14}; \textit{f}) \citet{Izo12} 

\end{table*}

The second galaxy sample is that of \citet{Eng08}. 
This sample was assembled to study 
metallicity effects on dust properties in starbursting galaxies so it is 
useful for our purposes as it spans a large metallicity range. We have 
added 31 objects from their study (those with 
oxygen abundances and H {\sc i} measurements, and which are not already 
present in the XMP sample). They form what will be called 
hereafter the Engelbracht sample. The relevant data for that sample are 
given in Table \ref{tab4}. It spans the metallicity range 
7.20$\leq$[O/H]$\leq$8.76.

\setcounter{table}{3}
\begin{table*}
 \centering
  \caption{Engelbracht Objects}
  	\label{tab4}
  \begin{tabular}{@{}lcccccccc@{}}
  \hline
  \multicolumn{1}{c}{Object} & $D$(Mpc)$^{h}$ & $M$(H {\sc i}) (M$_{\odot}$)$^{h}$ & $m_{g}$ & $M_{g}$ & log $M$(H {\sc i})/$L_{g}$ & [O/H] & log $M_{*}$(M$_{\odot}$) & log SFR(M$_{\odot}$ yr$^{-1}$)
\\
\multicolumn{1}{c}{(1)} & (2) & (3) & (4) & (5) & (6) & (7) & (8) & (9)
\\

\hline 
I Zw 18 & 13.9 & 1.25E+8 & 15.93$^{a}$ & -14.79 & 0.133 & 7.17$^{f}$  & 6.42$^{f}$ & -1.36$^{a}$
\\
UGC 4483 & 5.00 & 6.15E+7 & 15.00$^{(a,d)}$ & -13.49 & 0.345 & 7.46$^{f}$ & 5.16$^{f}$& -2.07$^{a}$
\\
ESO 146-G14 & 21.3 & 1.44E+9 & 14.94$^{(b,e)}$ & -16.70 & 0.430 & 7.66$^{h}$ & ... & ...
\\
DDO 187 & 2.07 & 1.05E+7 & 14.15$^{(b,b)}$ & -12.43 & 0.001 & 7.79$^{h}$ & ... & ... 
\\
Mrk 178 & 4.2 & 9.67E+6 & 14.03$^{a}$ & -14.09 & -0.699 & 7.88$^{h}$ & ... & -1.46$^{g}$
\\
UM 462 & 13.2 & 2.11E+8 & 14.32$^{b}$ & -16.28 & -0.236 & 7.95$^{h}$ & ... & -0.248$^{h}$
\\
UGC 4393 & 32.0 & 3.92E+9 & 13.12$^{b}$ & -19.41 & -0.219 & 7.95$^{h}$ & ... & -0.890$^{h}$
\\
Mrk 1450 & 17.8 & 3.29E+7 & 15.63$^{a}$ & -15.62 & -0.779 & 7.84$^{f}$ & 7.55$^{f}$ & -0.890$^{a}$
\\
UM 448 & 76.9 & 6.18E+9 & 14.47$^{b}$ & -19.96 & -0.241 & 8.32$^{f}$ & 10.45$^{f}$ & ...
\\
Mrk 170 & 18.4 & 3.88E+8 & 14.79$^{b}$ & -16.53 & -0.071 & 8.09$^{h}$ & ... & ...
\\
NGC 1569 & 1.45 & 2.45E+7 & 11.46$^{(b,b)}$ & -14.35 & -0.399 & 8.13$^{h}$ & ... & -1.52$^{h}$
\\
Mrk 1094 & 37.4 & 2.07E+9 & 14.09$^{a}$ & -18.77 & -0.240 & 8.15$^{h}$ & ... & -0.90$^{a}$
\\
NGC 3310 & 18.0 & 3.78E+9 & 10.82$^{b}$ & -20.46 & -0.655 & 8.18$^{h}$ & ... & 0.693$^{h}$
\\
NGC 1156 & 6.92 & 5.69E+8 & 11.89$^{(b,d)}$ &  -17.31 & -0.217 & 8.19$^{h}$ & ... & -0.904$^{h}$
\\
NGC 5253 & 2.79 & 6.11E+7 & 10.85$^{(b,d)}$ &  -16.38 & -0.814 & 8.19$^{h}$ & ... & -0.833$^{h}$
\\
NGC 4449 & 3.23 & 2.50E+8 & 9.97$^{(b,d)}$  & -17.58 & -0.682 & 8.23$^{h}$ & ... & -0.698$^{h}$
\\
II Zw 40 & 10.4 & 3.61E+8 & 11.75$^{a}$ &  -18.34 & -0.826 & 8.23$^{f}$ & 8.00$^{f}$ & 0.200$^{a}$
\\
NGC 7714 & 39.1 & 5.49E+9 & 12.74$^{(b,d)}$ & -20.22 & -0.396 & 8.26$^{h}$ & ... & 0.610$^{h}$
\\
NGC 1510 & 10.0 & 1.60E+9 & 13.27$^{(b,b)}$ & -16.73 & 0.464 & 8.33$^{h}$ & ... & ...
\\
NGC 4214 & 3.43 & 5.36E+8 & 10.02$^{(b,b)}$ & -17.66 & -0.383 & 8.36$^{h}$ & ... &  -1.04$^{h}$
\\
NGC 4670 & 21.0 & 1.15E+9 & 12.75$^{b}$ & -18.86 & -0.531 & 8.38$^{h}$ & ... & -0.037$^{h}$
\\
NGC 1140 & 19.4 & 2.37E+9 & 12.77$^{(b,b)}$ & -18.67 & -0.141 & 8.38$^{f}$ & 8.72$^{f}$ & ...
\\
NGC 2537 & 7.66 & 2.13E+8 & 11.94$^{b}$ & -17.48 & -0.712 & 8.44$^{h}$ & ... & -0.860$^{h}$
\\
NGC 3628 & 7.9 & 3.24E+9 & 10.11$^{(b,b)}$ & -19.38 & -0.289 & 8.57$^{h}$ & ... & -0.950$^{h}$
\\
NGC 2782 & 37.6 & 3.85E+9 & 11.75$^{b}$ & -21.13 & -0.915 & 8.59$^{h}$ & ... & ...
\\
NGC 5236 & 4.43 & 1.14E+10 & 8.04$^{(b,b)}$ & -20.19 & -0.067 & 8.62$^{h}$ & ... & 0.146$^{h}$
\\
NGC 3367 & 43.4 & 5.78E+9 & 11.91$^{b}$ & -21.28 & -0.798 & 8.62$^{h}$ & ... & ...
\\
NGC 4194 & 38.8 & 2.25E+9 & 12.93$^{b}$ & -20.01 & -0.700 & 8.67$^{h}$ & ... & 0.571$^{h}$
\\
NGC 2146 & 16.9 & 3.76E+9 & 11.13$^{(c,b)}$ & -20.01 & -0.477 & 8.68$^{h}$ & ... & ...
\\
NGC 2903 & 6.63 & 1.68E+9 & 9.37$^{(b,b)}$ & -19.74 & -0.719 & 8.68$^{h}$ & ... & ...
\\
Mrk 331 & 77.5 & 7.61E+9 & 14.61$^{b}$ & -19.84 & -0.103 & 8.76$^{h}$ & ... & ... \\
\hline
\end{tabular} 
\\
\noindent \noindent Notes: The columns are as follows. (1): Source name. (2): Distance. (3): H {\sc i} gas mass. (4): Apparent \I{g}-band magnitude derived using the $B-V$ color transformation as described in the text. Objects with two superscripts have both $B$ and $V$ magnitudes. Objects with one superscript have only $B$ magnitudes. (5): Absolute \I{g}-band magnitude. (6): Logarithm of the ratio of 
H {\sc i} gas mass to \I{g}-band luminosity. (7): Oxygen abundance 12+log O/H. (8): Logarithm of the stellar mass. (9): Logarithm of the star formation rate. Superscripts refer to the following sources:  \I{a}) \citet*{Gil03}; \I{b}) \citet{deVac}; \I{c}) \citet{Her96}; \I{d}) \citet{Tay05}; \I{e}) \citet{Zac06}; \I{f}) \citet{Izo14}; \I{g}) \citet{Ken08}; \I{h}) \citet{Eng08}
\\
\end{table*}

Finally, we have added a third galaxy sample, assembled from the
 {\it SIRTF} Nearby Galaxy Sample (SINGS) of \citet{Ken03}, hereafter called  
the SINGS sample. This sample is useful for 
increasing the number of galaxies and hence the statistics 
in the high mass and metallicity ranges. Selecting the galaxies with available 
oxygen abundances and H {\sc i} fluxes results in a sample of 53 galaxies.
The relevant data for the SINGS galaxies is given in Table \ref{tab5}.

The majority of the galaxies in the Engelbracht and SINGS samples do not 
possess $m_g$ but $m_B$. To convert $m_B$ into $m_g$, we have used 
the following transformation equation from \citet{Jes05}:

\begin{equation}
g = V + 0.74(B -V) -0.07.
\end{equation}
For galaxies that do not have a known $B-V$, we have adopted the mean $B-V$
given by \citet{But94} corresponding to the morphological type of the galaxy
as listed by \citet{Ken03}. 

\setcounter{table}{4}
\begin{table*}
 \centering
  \caption{SINGS Objects}
  	\label{tab5}
  \begin{tabular}{@{}lcccccccc@{}}
  \hline
  \multicolumn{1}{c}{Object} & $D$(Mpc)$^{f}$ & $M$(H {\sc i}) (M$_{\odot}$)$^{f}$ & $m_g$ & $M_g$ & log $M$(H {\sc i})/$L_g$ & [O/H]$^{b}$ & log $M_{*}$(M$_{\odot}$)$^{c}$ & log SFR(M$_{\odot}$ yr$^{-1}$)$^{e}$
\\
\multicolumn{1}{c}{(1)} & (2) & (3) & (4) & (5) & (6) & (7) & (8) & (9)
\\

\hline
DDO 154 & 3.66 & 1.90E+8 & 13.46$^{(a,a)}$ & -14.36 & 0.487 & 7.54 & 6.63 & -2.63
\\
DDO 53 & 2.42 & 2.76E+7 &  14.63$^{(a,a)}$ & -12.29 & 0.477 & 7.60 & 6.35 & -2.50
\\
Holmberg I & 4.74 & 1.92E+8 & 14.10$^{(a,d)}$ & -14.28 & 0.523 & 7.61 & 6.80 & -2.10
\\
DDO 165 & 3.01 & 4.88E+7 & 13.23$^{(a,a)}$ & -14.16 & -0.024 & 7.63 & 6.83 & -2.99
\\
Holmberg II & 4.89 & 1.24E+9 & 10.77$^{(a,a)}$ & -17.68 & -0.027 & 7.72 & 7.73 & -0.940
\\
NGC 5408 & 4.54 & 2.87E+8 & 11.60$^{(a,a)}$ & -16.69 & -0.266 & 7.81 & 8.29 & -1.02
\\
M81 Dw B & 8.25 & 6.10E+7 & 15.01$^{a}$ & -14.57 & -0.091 & 7.84 & 6.36 & -2.90
\\
IC 2574 & 3.09 & 8.72E+8 & 10.62$^{(a,a)}$ & -16.83 & 0.161 & 7.85 & 8.16 & -1.31
\\
NGC 2915 & 3.14 & 2.41E+8 & 12.79$^{(a,a)}$ & -14.69 & 0.458 & 7.94 & 7.57 & ... 
\\
NGC 1705 & 5.94 & 1.28E+8 & 12.63$^{(a,a)}$ & -16.24 & -0.434 & 7.96 & ... & -0.990
\\
NGC 1482 & 19.9 & 6.19E+8 & 12.81$^{(a,a)}$ & -18.68 & -0.728 & 8.11 & 9.99 & ...
\\
NGC 4631 & 6.93 & 7.29E+9 & 9.43$^{a}$ & -19.77 & -0.093 & 8.12 & 9.76 & 0.040
\\
Holmberg IX & 2.97 & 2.56E+7 & 14.17$^{(a,a)}$ & -13.19 & 0.084 & 8.14 & ... & -3.10
\\
NGC 4536 & 28.9 & 6.86E+9 & 10.67$^{(a,a)}$ & -21.63 & -0.864 & 8.21 & 9.49 & ...
\\
NGC 5713 & 29.5 & 1.05E+10 & 11.37$^{b}$ & -20.98 & -0.419 & 8.24 & 10.07 & ...
\\
NGC 925 & 9.98 & 6.02E+9 & 10.58$^{b}$ & -19.42 & -0.036 & 8.25 & 9.48 & -0.180
\\
NGC 24 & 7.36 & 9.43E+8 & 12.05$^{b}$ & -17.28 & 0.015 & 8.26 & ... & -1.22
\\
NGC 3621 & 6.92 & 9.46E+9 & 9.20$^{b}$ & -20.00 & -0.072 & 8.27 & 9.43 & 0.070
\\
NGC 4559 & 9.0 & 6.74E+9 & 10.44$^{b}$ & -19.33 & 0.049 & 8.29 & 8.93 & ...
\\
Mrk 33 & 24.2 & 5.95E+8 & 13.28$^{a}$ & -18.64 & -0.729 & 8.30 & ... & ...
\\
NGC 4736 & 4.40 & 6.00E+8 & 8.47$^{b}$ & -19.75 & -1.17 & 8.31 & 10.34 & -0.700
\\
NGC 5474 & 6.23 & 1.03E+9 & 11.31$^{b}$ & -17.66 & 0.796 & 8.31 & 8.70 & -1.07
\\
NGC 2403 & 4.47 & 8.76E+9 & 8.60$^{b}$ & -19.65 & 0.035 & 8.33 & ... & -0.400
\\
NGC 2798 & 27.0 & 2.33E+9 & 12.73$^{b}$ & -19.43 & -0.453 & 8.34 & 10.04 & ...
\\
NGC 3198 & 11.19 & 7.17E+9 & 10.04$^{b}$ & -20.20 & -0.272 & 8.34 & 9.85 & ...
\\
NGC 3627 & 6.12 & 3.93E+8 & 9.04$^{b}$ & -19.89 & -1.41 & 8.34 & 10.57 & -0.420
\\
NGC 7331 & 15.0 & 9.31E+9 & 9.70$^{b}$ & -21.18 & -0.551 & 8.34 & 10.58 & ...
\\
NGC 628 & 10.48 & 7.83E+9 & 9.80$^{a}$ & -20.30 & -0.274 & 8.35 & 9.57 & ...
\\
NGC 4625 & 9.58 & 1.07E+9 & 12.90$^{b}$ & -17.01 & 0.177 & 8.35 & 8.72 & -1.28
\\
NGC 4725 & 24.5 & 1.52E+10 & 10.48$^{b}$ & -21.47 & -0.454 & 8.35 & 10.58 & ...
\\
NGC 7552 & 20.6 & 4.09E+9 & 11.09$^{b}$ & -20.48 & -0.628 & 8.35 & ... & ...
\\
NGC 2976 & 2.18 & 5.09E+7 & 10.76$^{(a,a)}$ & -15.93 & -0.713 & 8.36 & 8.97 & -1.54
\\
NGC 5033 & 15.7 & 1.30E+10 & 10.29$^{a}$ & -20.69 & -0.210 & 8.37 & ... & ...
\\
NGC 3521 & 7.7 & 3.43E+9 & 8.90$^{b}$ & -20.53 & -0.725 & 8.39 & ... & -0.130
\\
NGC 5055 & 7.59 & 6.48E+9 & 9.12$^{b}$ & -20.28 & -0.348 & 8.40 & 10.76 & -0.220
\\
NGC 6946 & 5.52 & 3.64E+9 & 8.98$^{(a,a)}$ & -19.73 & -0.379 & 8.40 & 9.96 & 0.300
\\
NGC 3031 & 1.68 & 7.79E+8 & 7.70$^{a}$ & -18.43 & -0.528 & 8.43 & ... & ...
\\
NGC 3773 & 10.1 & 5.99E+7 & 13.44$^{b}$ & -16.58 & -0.903 & 8.43 & 8.31 & ...
\\
NGC 4254 & 36.3 & 2.38E+10 & 9.95$^{(a,a)}$ & -22.85 & -0.811 & 8.45 & 9.61 & ...
\\
NGC 1097 & 15.4 & 7.98E+9 & 9.77$^{(a,a)}$ & -21.17 & -0.614 & 8.47 & 10.74 & ...
\\
NGC 4321 & 13.1 & 2.00E+9 & 9.95$^{(a,a)}$ & -20.64 & -1.00 & 8.50 & 10.36 & ...
\\
NGC 3034 & 5.68 & 5.33E+9 & 9.06$^{a}$ & -19.71 & -0.205 & 8.51 & ... & ...
\\
NGC 3184  & 9.27 & 2.13E+9 & 10.25$^{a}$ & -19.59 & -0.556 & 8.51 & 9.24 & ...
\\
NGC 3049 & 21.9 & 1.31E+9 & 12.78$^{a}$ & -18.92 & -0.499 & 8.53 & 	8.58 & ...
\\
NGC 2841 & 11.52 & 4.68E+9 & 9.20$^{(b,a)}$ & -21.11 & -0.822 & 8.54 & 10.17 & ...
\\
NGC 4826 & 3.54 & 1.23E+8 & 8.86$^{(a,a)}$ & -18.89 & -1.51 & 8.54 & 9.99 & -1.10
\\
NGC 5194 & 8.33 & 2.77E+9 & 8.86$^{a}$ & -20.74 & -0.902 & 8.55 & ... & 0.220
\\
NGC 1512 & 9.84 & 5.88E+9 & 10.95$^{(a,a)}$ & -19.01 & 0.117 & 8.56 & 10.10 & -0.570
\\
NGC 3351 & 8.4 & 1.03E+9 & 10.23$^{b}$ & -19.39 & -0.791 & 8.60 & 10.28 & -0.660
\\
\hline
\end{tabular}
\\
\noindent \noindent Notes: The columns are as follows. (1): Source name. (2): Distance. (3): H {\sc i} gas mass. (4): Apparent \I{g}-band magnitude derived using the $B-V$ color transformation as described in the text. Objects with two superscripts have both $B$ and $V$ magnitudes. Objects with one superscript have only 
$B$ magnitudes. (5): Absolute \I{g}-band magnitude. (6): Logarithm of the ratio of H {\sc I} gas mass to \I{g}-band luminosity. (7): Oxygen abundance 12+log O/H. (8): Logarithm of the stellar mass. (9): Logarithm of the star formation rate. Superscripts refer to the following sources: \I{a}) \citet{deVac}; \I{b}) \citet{Mou10}; \I{c}) \citet{Ski11}; \I{d}) \citet{Mar99}; \I{e}) \citet{Ken08}; \I{f}) \citet{Ken03}
\\
\end{table*}
 
To compare the four samples with one another,
we have attempted to minimize as much as possible systematic effects 
by using quantities that are derived in the same manner 
in the different samples. For example, the inclinations for the dwarf galaxies
are derived using the same intrinsic ratio $r_0$ in both the GBT and XMP
samples. The metallicities of many objects in the XMP sample came from 
publications from our group, ensuring that they are derived in the same way 
as the metallicities of the GBT sample. In any case, we expect small 
systematic differences in quantities such as stellar masses and metallicities
to be overcome by the wide range of parameter space covered by these samples.   Combining the GBT, XMP, Engelbracht and SINGS samples, we end up with a
final sample of 151 objects. All data are scaled, when necessary, to a distance based on $H_{0}$ = 73.0 km s$^{-1}$ Mpc$^{-1}$.
  
Fig. \ref{fig3} shows the metallicity histogram of the four samples. 
Examination of the figure shows that including the comparison samples 
extend the oxygen abundance 
range from 7.1 (1/40 solar) to 8.7 (solar),
with the XMP and GBT low-mass low-luminosity galaxies 
 covering the low-metallicity range, 
while the Engelbracht and SINGS higher-mass higher-luminosity galaxies    
span the high-metallicity range. 
Extending the metallicity range of the total sample extends the 
ranges of $M_g$ (from $-$9 to $-$23 mag, Fig. \ref{fig6}), and of SFRs (Fig. \ref{fig9}) and 
sSFRs (Fig. \ref{fig10}). 

\section{Analysis}

We now combine the four galaxy samples to explore possible correlations 
between various quantities.  

\subsection{H {\sc i} mass-to-optical light ratio as a function of metallicity and absolute magnitude} 
We study here how the H~{\sc i} gas content depends on the galaxy's 
metallicity and luminosity.
We first plot in Fig. \ref{fig4} the  $M$(H~{\sc i})/$L_g$ ratio against oxygen 
abundance for the galaxies in the four samples. 
The figure shows a clear trend of increasing ratios 
for lower metallicity galaxies. A linear least-square fit to the data, where [O/H] = 12+log O/H,  
gives:

\begin{equation}
 \log [M({\rm HI})/L_{g}] = (-0.83\pm0.08) \times [{\rm O/H}] + (6.47\pm0.66).
\end{equation} 

\noindent
\citet*{Sta92} found the $M$(H~{\sc i})/$L_g$ ratio to increase  
with decreasing galaxy luminosity.
A linear least-square fit to our data 
gives (Fig. \ref{fig5}):

\begin{equation}
 \log[M({\rm HI})/L_{g}] = (-0.30\pm0.03) \times \log L_g + (2.53\pm0.29).
\end{equation}

\noindent
The slope -0.30 of this relation is the same    
as the one obtained by \citet{Sta92}, based on a smaller sample of 
36 BCDs and low-surface-brightness 
dwarf galaxies, and spanning only 10 magnitudes in $M_B$ (from $-$11 to $-$19 mag)
instead of our luminosity range of 14 magnitudes in $M_g$ (from $-$9 to $-$23 mag).
Our slope is however steeper than the one of $-$0.2 obtained by \citet{Lee02}. 
The difference probably arises because of the more restricted luminosity range
of the sample of \citet{Lee02} , spanning only 4 mag in  $M_B$, 
from $-$14 to $-$18 mag.   

\subsection{Metallicity-luminosity and Mass-metallicity relations}

Fig.\ref{fig6} shows the well-known metallicity-luminosity relation for galaxies.  
Our fit (solid line) 

\begin{equation}
 [{\rm O/H}] = (-0.12\pm0.01) \times M_g + (5.82\pm0.12)
\end{equation}

\noindent
is in good agreement, within the errors, with the ones obtained for a local dwarf irregular galaxy sample by \citet{Ski89} (dot-dashed line), and  for   
a local emission-line galaxy sample by 
\citet{Gus09} (dashed line). 
We note that four of the lowest-metallicity objects in our total sample, 
SBS 0335$-$052E, J0113+0052 = UGC 772, I Zw 18 and SBS 0335$-$052W  
(labeled in Fig. \ref{fig6}) deviate 
strongly from the above relation, being several magnitudes 
too bright for their oxygen abundance.  
These four galaxies are all undergoing strong bursts of star formation which 
increase significantly their luminosities.  
\citet{Izo11} have shown that all galaxies undergoing strong starbursts also 
define their own luminosity-metallicity relation. This relation would have about the same slope, but would be shifted by several magnitudes towards higher luminosities. Similarly, \citet{Fil13} found the metallicity-luminosity relation for their XMP galaxy sample to be shifted towards higher luminosities.

Fig. \ref{fig7} shows the relation between stellar mass and metallicity. 
The best fit (solid line) is given by the equation:
  
\begin{equation}
[{\rm O/H}] = (6.00\pm0.11) + (0.23\pm0.01) \times \log M_*.
\end{equation} 
\noindent
where $M_*$ represents the stellar mass in units of solar masses.

For comparison, we have also shown the relations derived by \citet{Tre04} 
(dashed line) and \citet{Lee06} (dotted line). It is seen that, in the 
region of stellar mass overlap (8.6$\leq$log $M_*$/M$_\odot$$\leq$11), the 
\citet{Tre04} curve is  
offset towards higher metallicities relative to ours by 0.2-0.7 dex. 
In this stellar mass range, our fit is mainly defined by the SINGS data 
points. The oxygen abundances of the SINGS galaxies are determined 
by \citet{Mou10}, using both a theoretical \citep{Kob04} and an empirical 
\citep{Pil05} strong-line abundance 
calibration. We have adopted 
here the abundances resulting from the empirical calibration. 
The shift of the \citet{Tre04} curve is probably due to 
systematically too high oxygen abundances derived for their SDSS objects.  
Our fit is slightly flatter than the linear least-square fit
[O/H] = (5.65$\pm$0.23) + (0.30$\pm$0.03) $\times$ $\log M_*$  obtained by  \citet{Lee06}. They are however consistent within 2$\sigma$.  

\begin{figure*}
  \centering
\hbox{
  \includegraphics[width=.45\linewidth]{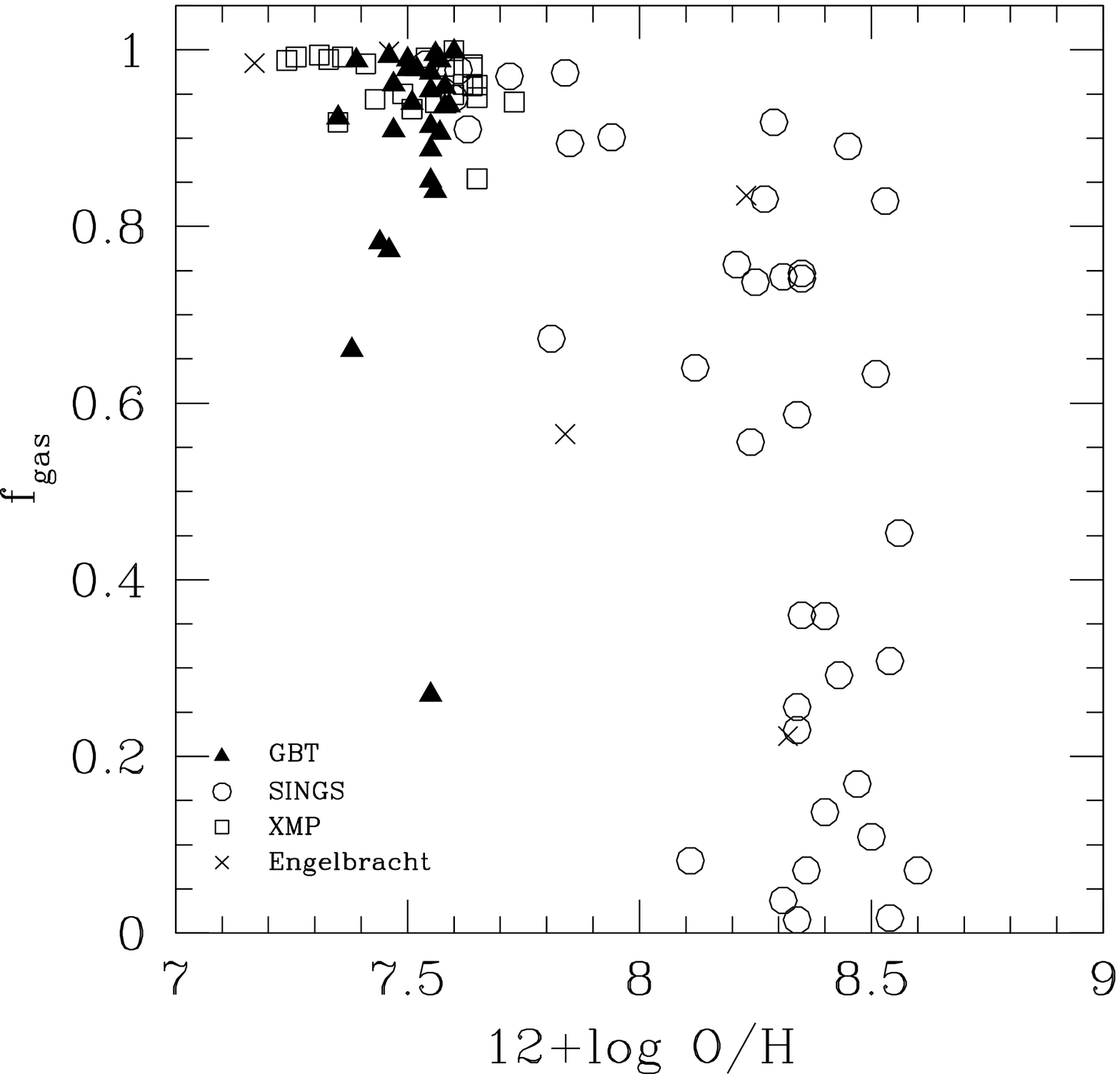}
  \hspace{0.3cm}\includegraphics[width=.45\linewidth]{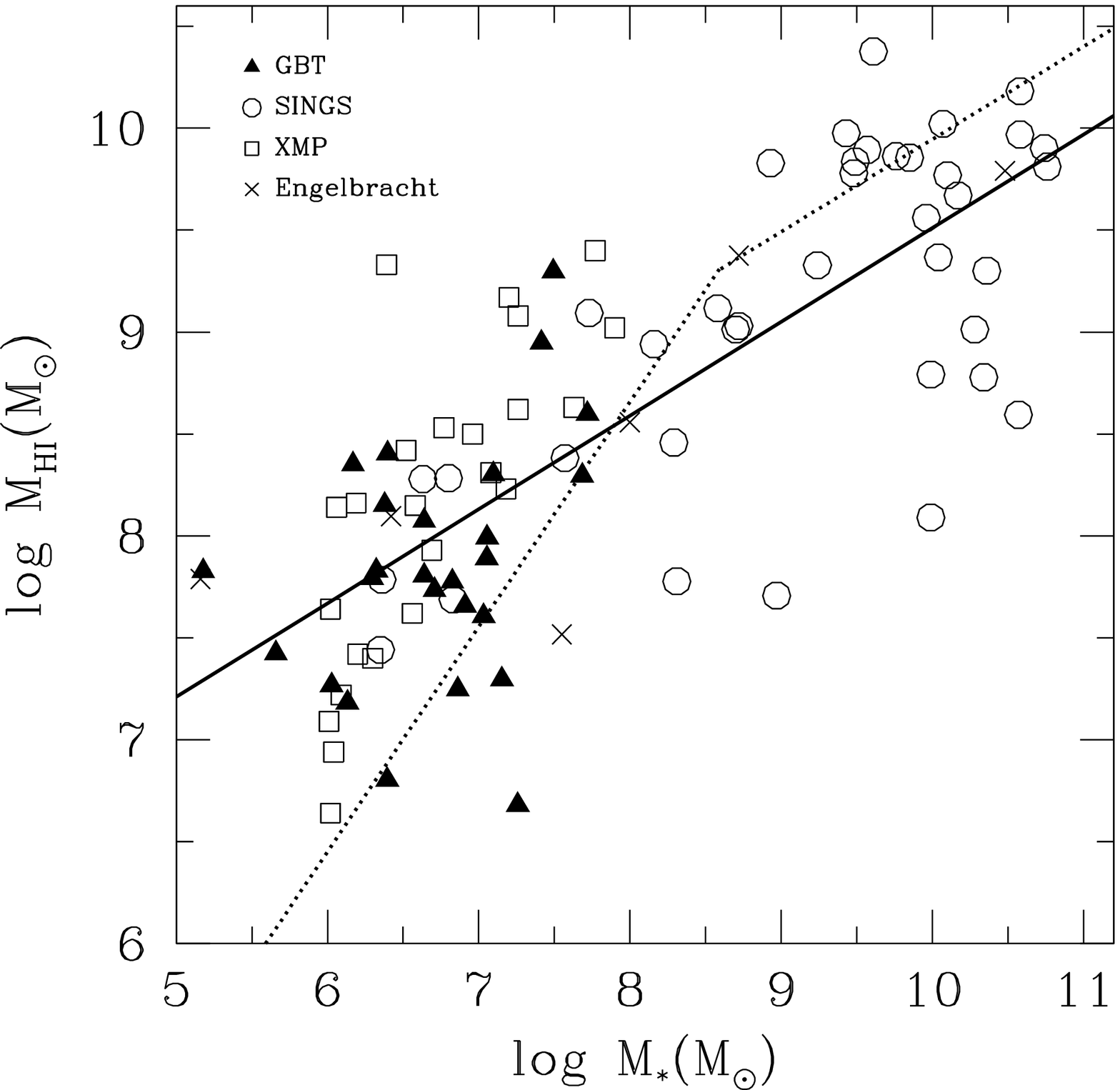}}
\caption{Plots of: (left) gas mass fraction versus metallicity; (right) neutral hydrogen mass versus 
stellar mass. The solid line shows the linear least-square fit to all 
galaxies in the GBT and the 3 comparison samples. It is described 
by the relation
log $M$(H {\sc i}) = (0.46$\pm$0.04) $\times$ log $M_*$ + (4.91$\pm$0.31).
The dashed line shows the fit
obtained by \citet*{Bra15} for their larger SDSS Data Release 8 galaxy sample.}
\label{fig8}
\end{figure*}

\begin{figure*}
  \centering
\hbox{
  \includegraphics[width=.45\linewidth]{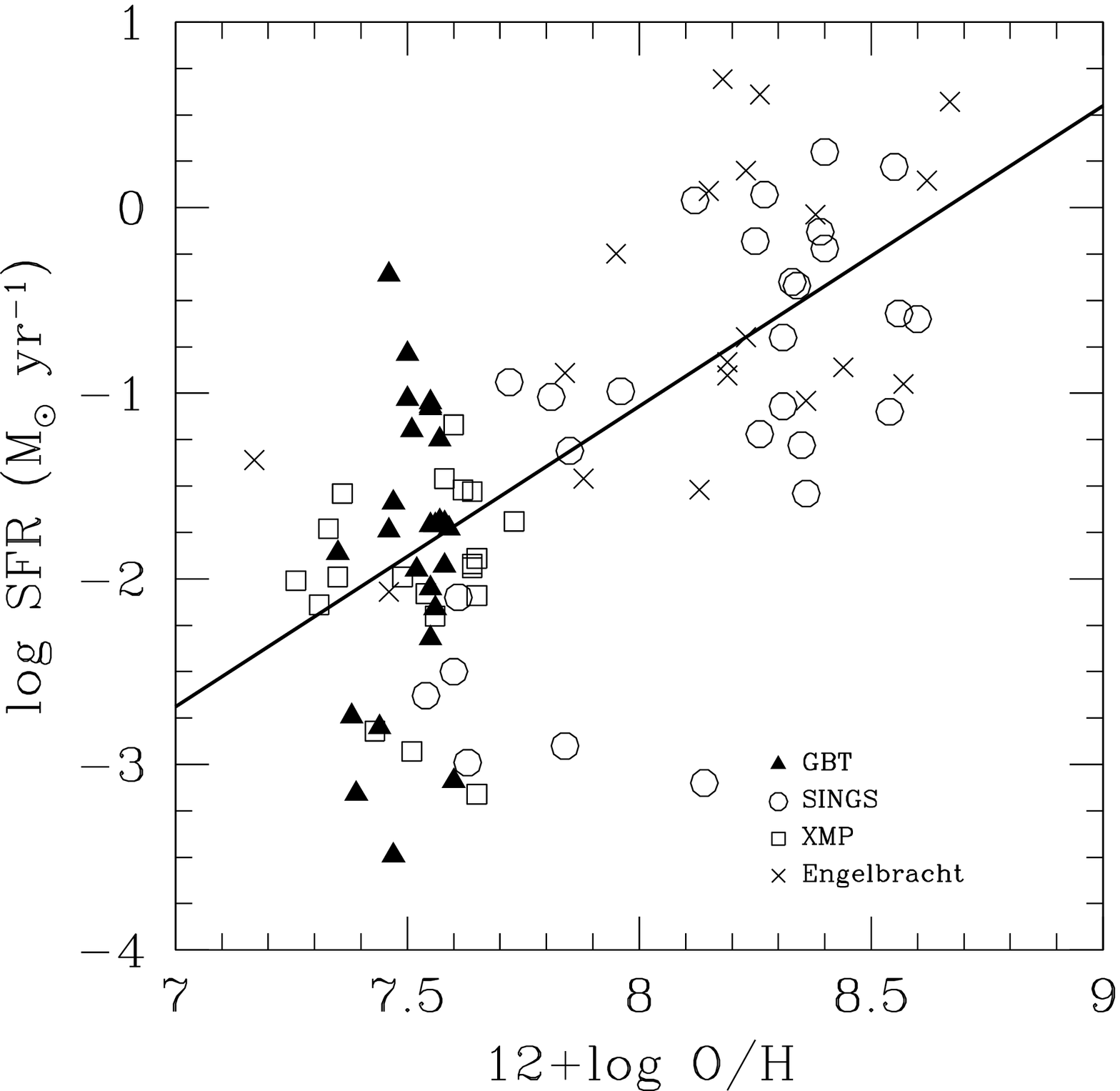}
  \hspace{0.3cm}\includegraphics[width=.45\linewidth]{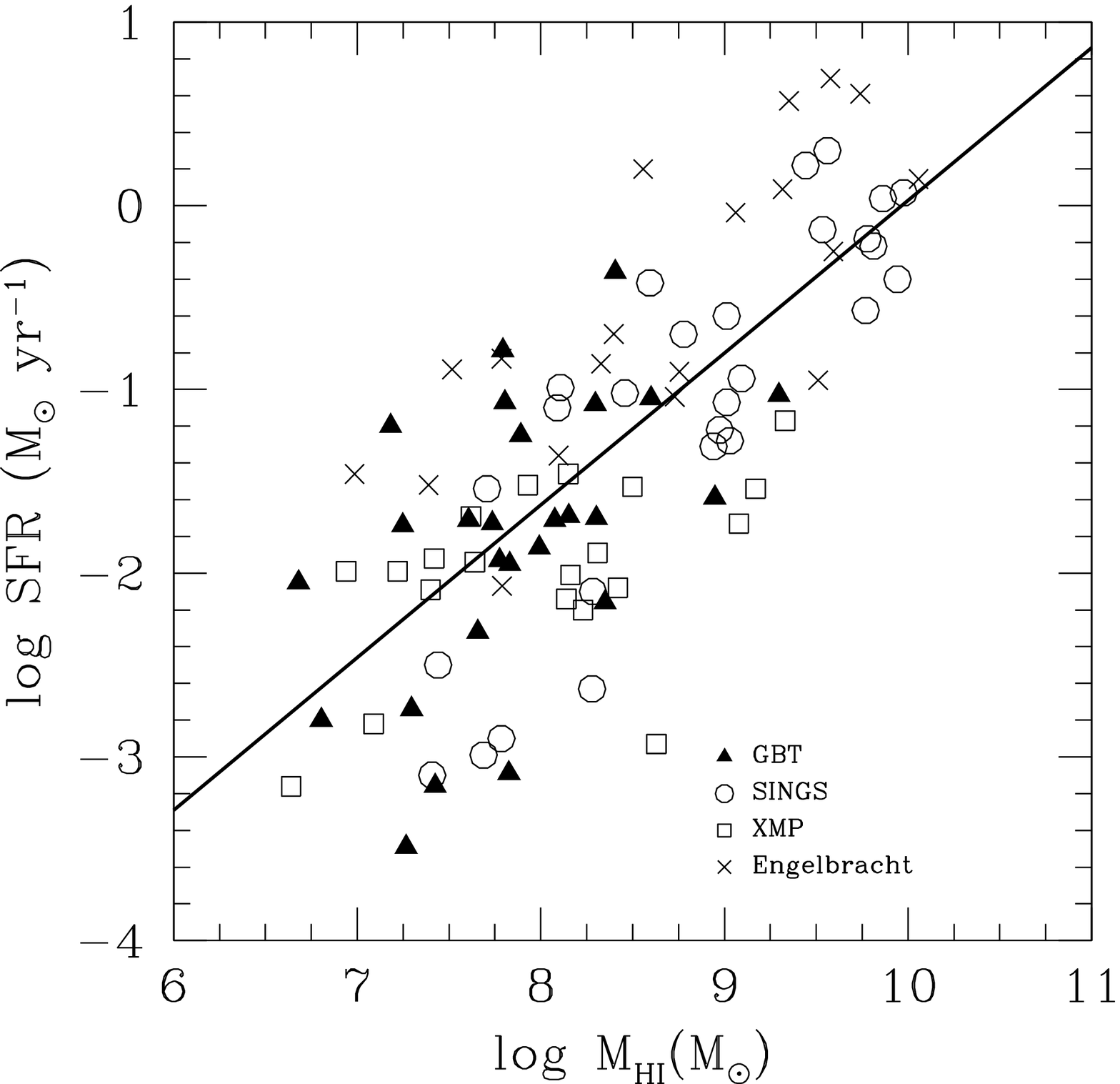}}
\caption{Plots of SFR versus: (left) metallicity. The solid line shows the least-square fit to all galaxies, described by the relation log SFR = (1.61$\pm$0.19) $\times$ [O/H] $-$ (14.03$\pm$1.50); 
(right) neutral hydrogen mass. The least-square fit is given by 
log SFR = (0.83$\pm$0.08) $\times$ log $M$(H~{\sc i}) $-$ (8.27$\pm$0.67).}
\label{fig9}
\end{figure*}

\begin{figure*}
  \centering
\hbox{
  \includegraphics[width=.45\linewidth]{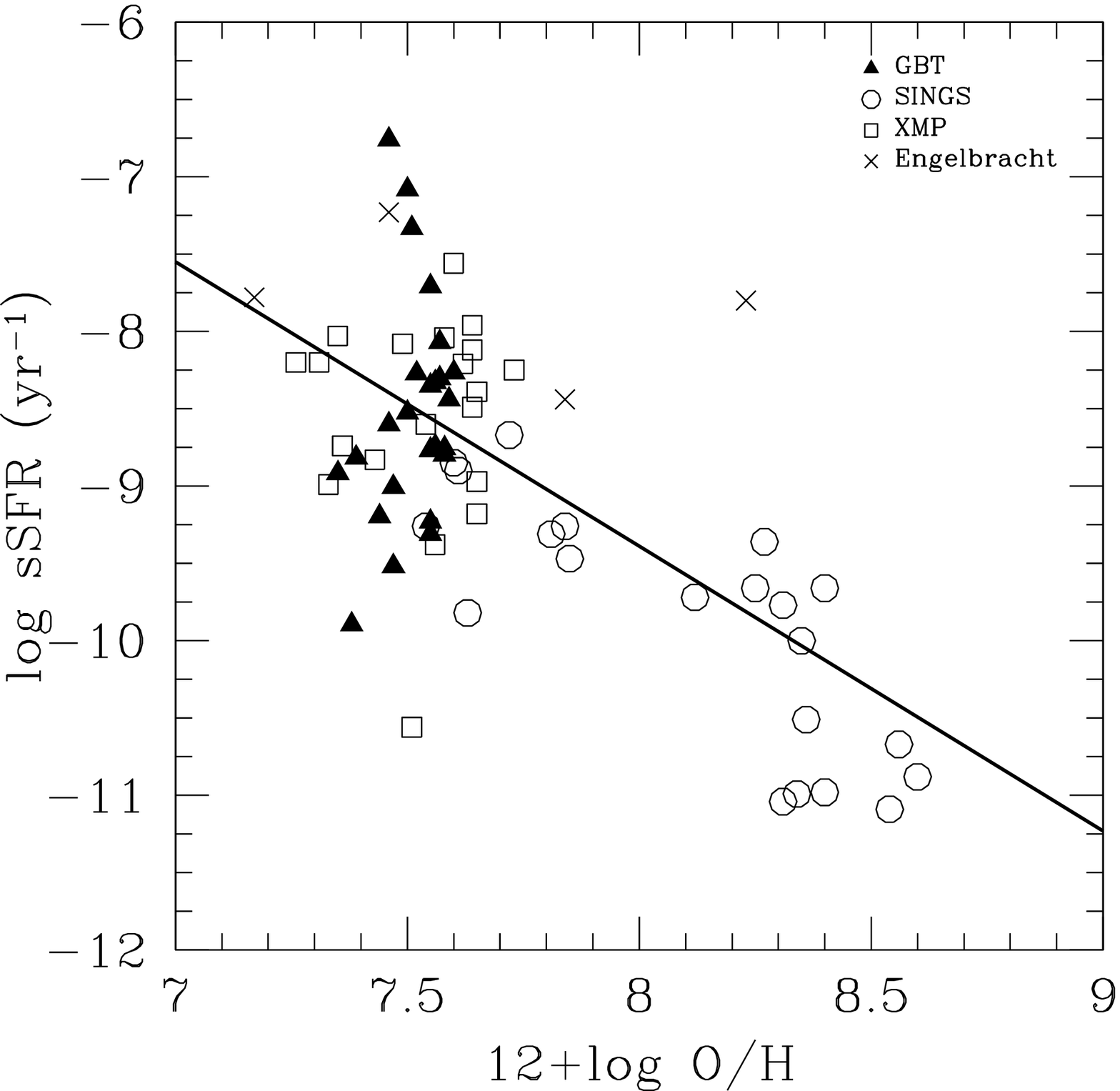}
\centering
  \hspace{0.3cm}\includegraphics[width=.45\linewidth]{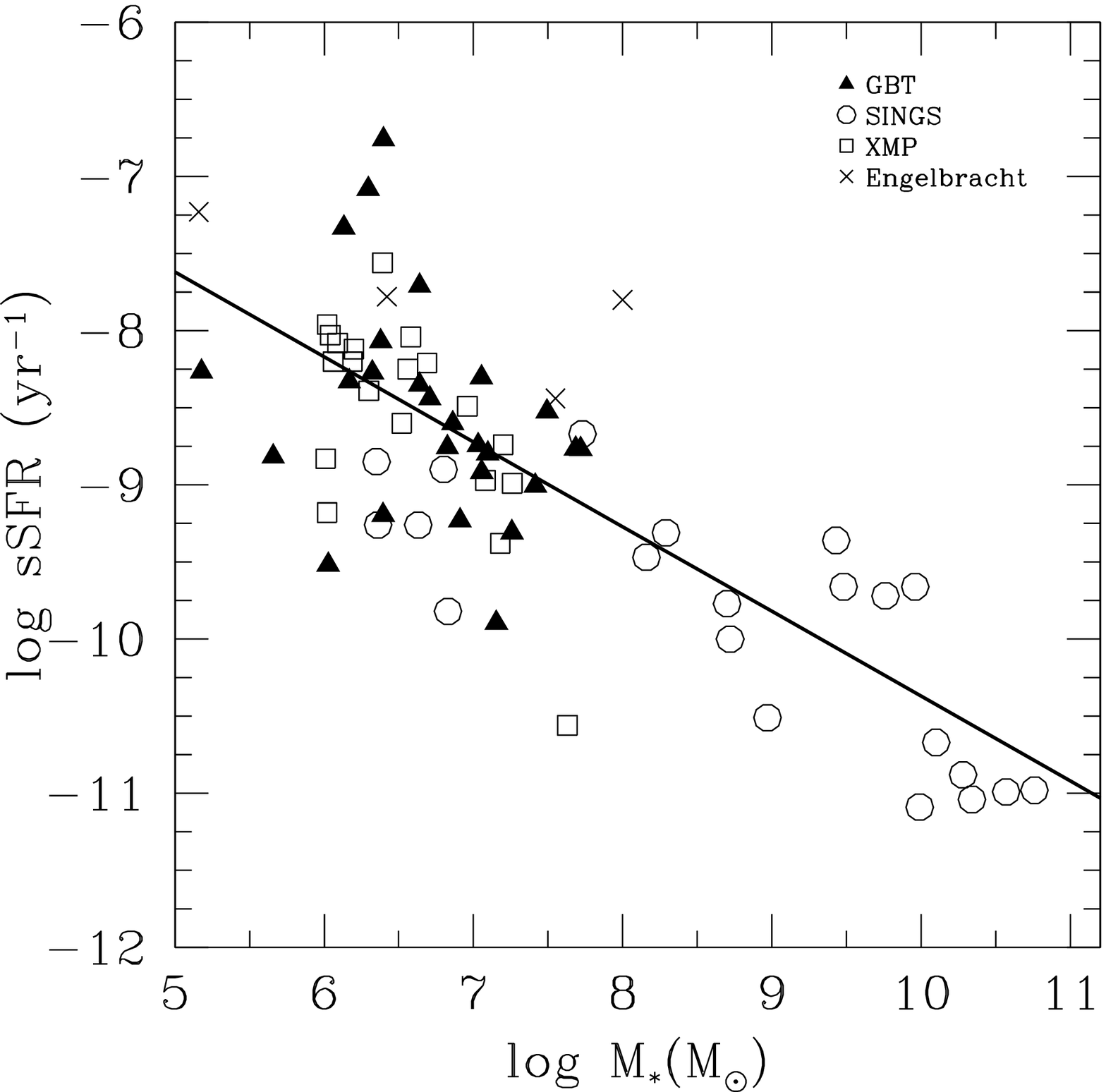}
}
\caption{Plots of the specific star formation rate sSFR versus: (left) metallicity. The solid line shows the least-square fit to all galaxies, described by the relation log sSFR = ($-$1.85$\pm$0.25) $\times$ [O/H] + (5.33$\pm$1.89);
(right) stellar mass. The least-square fit is given by 
log sSFR = ($-$0.55$\pm$0.05) $\times$ log $M_{*}$ $-$ (4.87$\pm$0.40).} 
\label{fig10}
\end{figure*}

\begin{figure*}
  \centering
\hbox{
  \includegraphics[width=.45\linewidth]{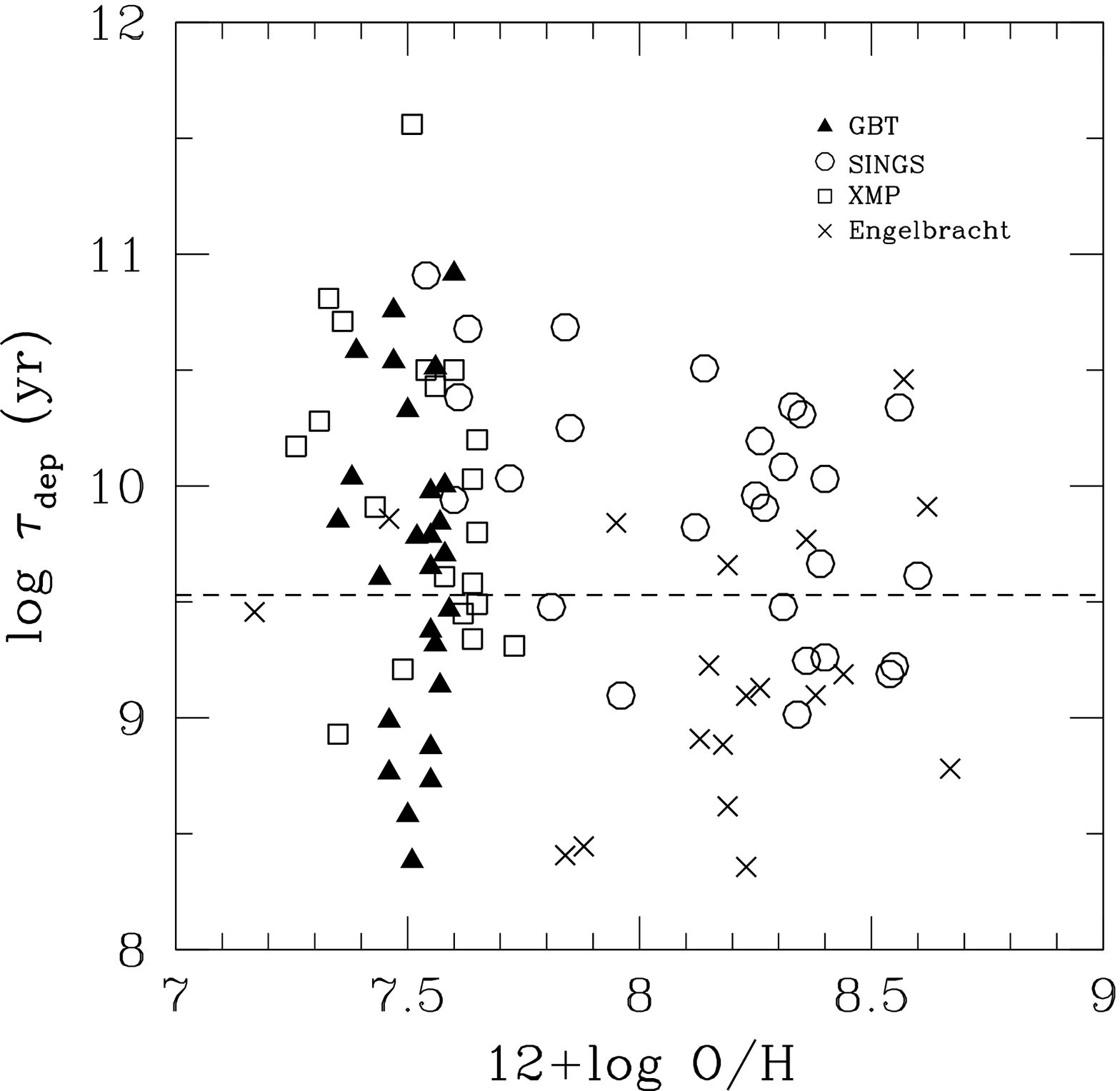}
\centering
\hspace{0.3cm}\includegraphics[width=.45\linewidth]{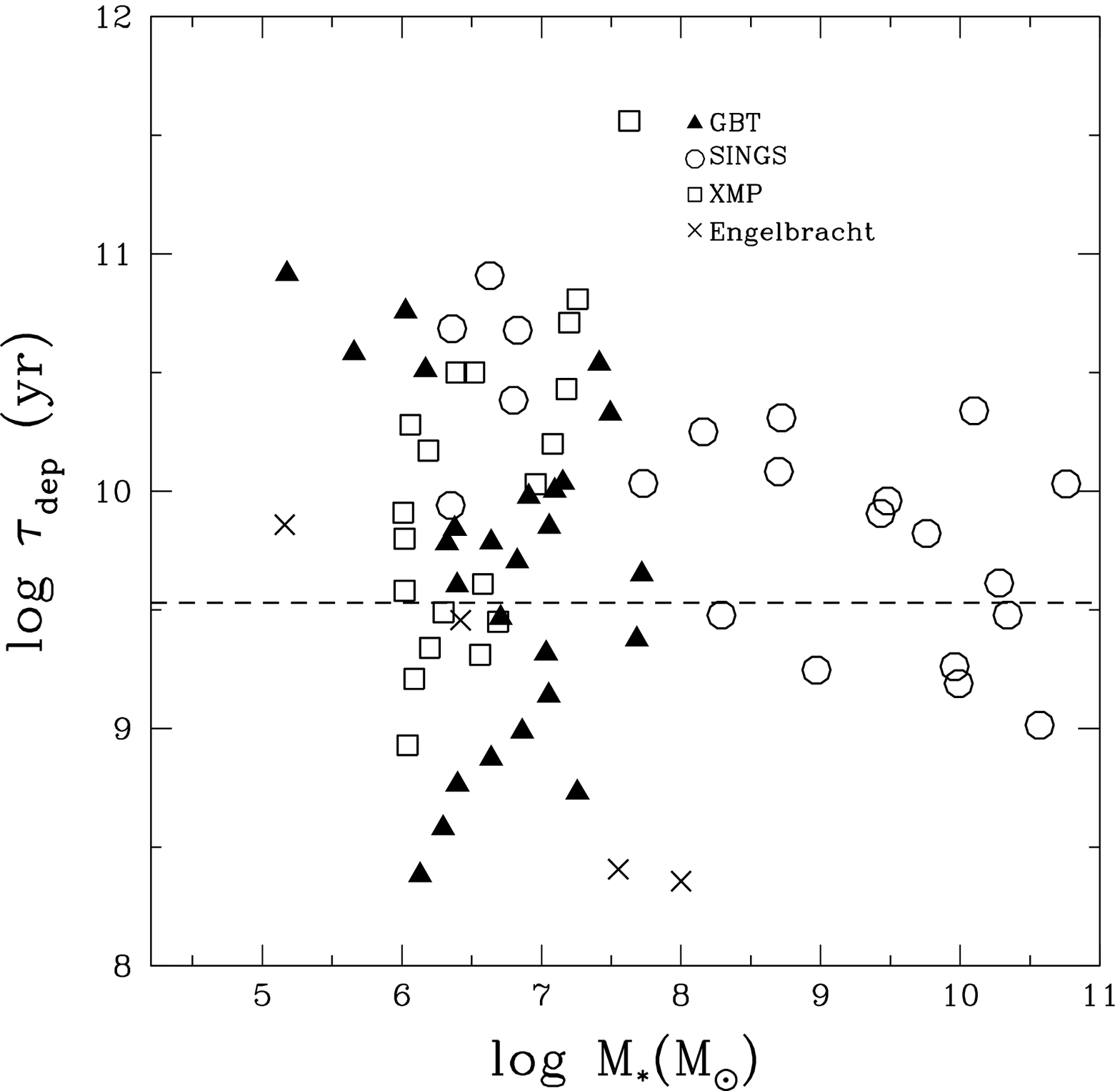}}
\caption{Plots of the gas depletion time versus (left) oxygen abundance, 
 and (right) stellar mass,
  for all galaxies in the GBT and the 3 comparison samples.
The dashed line in both plots 
shows the constant $\tau_{dep}$ = 3.4 Gyr found by \citet{Sch10}
for their {\it GALEX} Arecibo SDSS galaxy sample.
}
\label{fig11}
\end{figure*}

\subsection{The gas mass fraction as a function of metallicity}

In Fig. \ref{fig8} (left), we plot the distance-independent 
gas mass fraction (Eq. \ref{gmf}) as a function of the oxygen abundance. 
It is seen that while the high-metallicity ([O/H] $\geq$ 8.0) 
SINGS galaxies span the range 
0$\leq$$f_{gas}$$\leq$0.95, the lower-metallicity GBT objects (with the exception of one galaxy, the 
BCD J1214+0940 with $f_{gas}$= 0.27) span 
the more restricted range 0.65$\leq$$f_{gas}$$\leq$0.99, i.e. the vast majority of  
low-metallicity ([O/H] $\leq$ 7.8) have more than 65\% of their 
baryonic mass in gaseous form. This is consistent with the $f_{gas}$ range 0.3--0.99 
found by \citet{Bra15} for their sample of 148 isolated low-mass galaxies. 
As noted by those authors, the lower limit of $\sim$0.3 for $f_{gas}$ puts 
constraints on the internal feedback processes in low-mass and low-metallicity
 galaxies as they should not remove all of the galaxy's atomic gas. We note 
also that in the low-metallicity regime,
 the GBT sample spans a larger range of gas mass fractions than the XMP sample of \citet{Fil13}: the vast majority of XMP galaxies have $f_{gas}$ $\geq$0.85.   
We have examined J1214+0940 to  
see if it possesses any particular characteristic that 
would make it relatively gas-poor, but found none.    
The median $f_{gas}$ for the GBT sample is 0.94, while 
the mean $f_{gas}$ is equal to 0.90$\pm$0.15. Our median value is higher than the value $f_{gas}$= 0.82$\pm$0.13 obtained by \citet{Bra15} for their larger SDSS dwarf galaxy sample. Evidently, the low-metallicity selection criterion selects out more gas-rich dwarf galaxies. 

In Fig. \ref{fig8} (right), we have plotted the atomic gas mass against the stellar mass. 
While our data for the GBT and the 3 comparison samples 
can be fitted by a single linear least-square 
fit (solid line), \citet{Bra15} obtained for their considerably larger sample 
of SDSS dwarf and non-dwarf galaxies a fit with a 
similar slope for non-dwarf galaxies ($M_*$$\geq$10$^{8.6}$ M$_\odot$) and a 
steeper slope for dwarf galaxies ($M_*$$\leq$10$^{8.6}$ M$_\odot$), with 
a break in the fit at about 10$^{8.6}$ M$_\odot$ (dashed line).
We are not able to see the slope break in our data 
because of the smaller number of galaxies with $M_*$$\geq$10$^{8.6}$ M$_\odot$ 
in our sample.

 

\subsection{Star formation rate}

Fig. \ref{fig9} (left) and Fig. \ref{fig9} (right) show respectively 
the trends of SFR with oxygen abundance and H {\sc i} mass. 
It is seen that 
there is a general correlation of increasing SFR for galaxies 
with higher metallicities (and hence higher stellar masses) and 
higher H {\sc i} masses.
However, when the 
SFR is normalized to the galaxy's stellar mass, i.e. if we plot the specific 
sSFR = SFR/$M_*$ against metallicities (Fig. \ref{fig10}, left) and stellar masses 
(Fig. \ref{fig10}, right), 
then the trend is reversed. The galaxies with the lowest metallicities and 
stellar masses 
have the highest sSFRs. Similar trends were found by \citet{Fil13} for 
their XMP galaxy sample and by 
\citet{Hun15} for 
a sample of metal-poor BCDs together with other galaxy samples from the 
literature. 




\subsection{Depletion time scales}

We consider here the H {\sc i} depletion time 
scale as defined by $\tau_{dep}$ (yr) = $M$(H {\sc i})/SFR. This quantity measures 
the time left for a galaxy to form stars at the present rate before 
exhausting its gas supply. Galaxies with $\tau_{dep}$ less than the Hubble time 
cannot make stars at the present rate 
without running out of fuel.

Fig. \ref{fig11} (left) and Fig. \ref{fig11} (right) show that there is no dependence of    
log $\tau_{dep}$ on either oxygen abundance or M$_*$, with a large  
scatter. This is in agreement 
with the conclusion of \citet{Sch10} who find a relatively constant mean $\tau_{dep}$ = 3.4 Gyr across their {\it GALEX} Arecibo SDSS galaxy sample.
\citet{Hun15} also find a similar mean constant value, also 
with a large scatter \citep[see also ][]{Fil16}. 
This mean value is shown 
by a dashed horizontal line in both panels of Fig. \ref{fig11}. While this line 
bissects well our data, the scatter about the mean is  
$\sim$1.5 dex on either side of the line. Using interferometric 
maps of a sample of spiral and dwarf irregular galaxies, 
\citet{Roy15} have also found that 
the  H {\sc i} depletion time scale shows 
no strong dependence on metallicity within individual galaxies.  

Why does sSFR= SFR/$M_*$ decrease steeply with increasing $M_*$ (Fig. \ref{fig10}, right) while  $\tau_{dep}$ or, equivalently, the inverse ratio which 
represents  
the H {\sc i}-based star formation efficiency  
SFE = SFR/$M$(H {\sc i}) remains relatively constant with $M_*$ (Fig. \ref{fig11}, right)? \citet{Sch10} interpret 
this approximate constancy as indicating that external processes or feedback mechanisms controlling
 the gas supply are important for regulating star formation in 
massive galaxies. Our results show that these regulation mechanisms operate 
also in low-mass galaxies.  \citet{San14,San15} 
have argued, based on observations of metallicity drops shown by 
localized starburts in XMD galaxies, that accretion flows of external 
metal-poor gas may be a dominant regulation mechanism.     
        
The majority 
of the GBT galaxies (77\%) have $\tau_{dep}$ shorter than
the Hubble time. This is consistent with the idea that their star formation histories are composed 
of short bursts lasting $\sim$10$^{7-8}$ yr, interspersed with quiescent 
periods of $\sim$10$^9$ yr \citep{Thu08}. 
This appears also to be the case for the Engelbracht galaxies 
classified as BCDs. 
As for the XMP and SINGS galaxies,  
slightly more than half have $\tau_{dep}$ shorter than the Hubble time.      


\begin{figure}
  \centering
  \includegraphics[width=.8\linewidth]{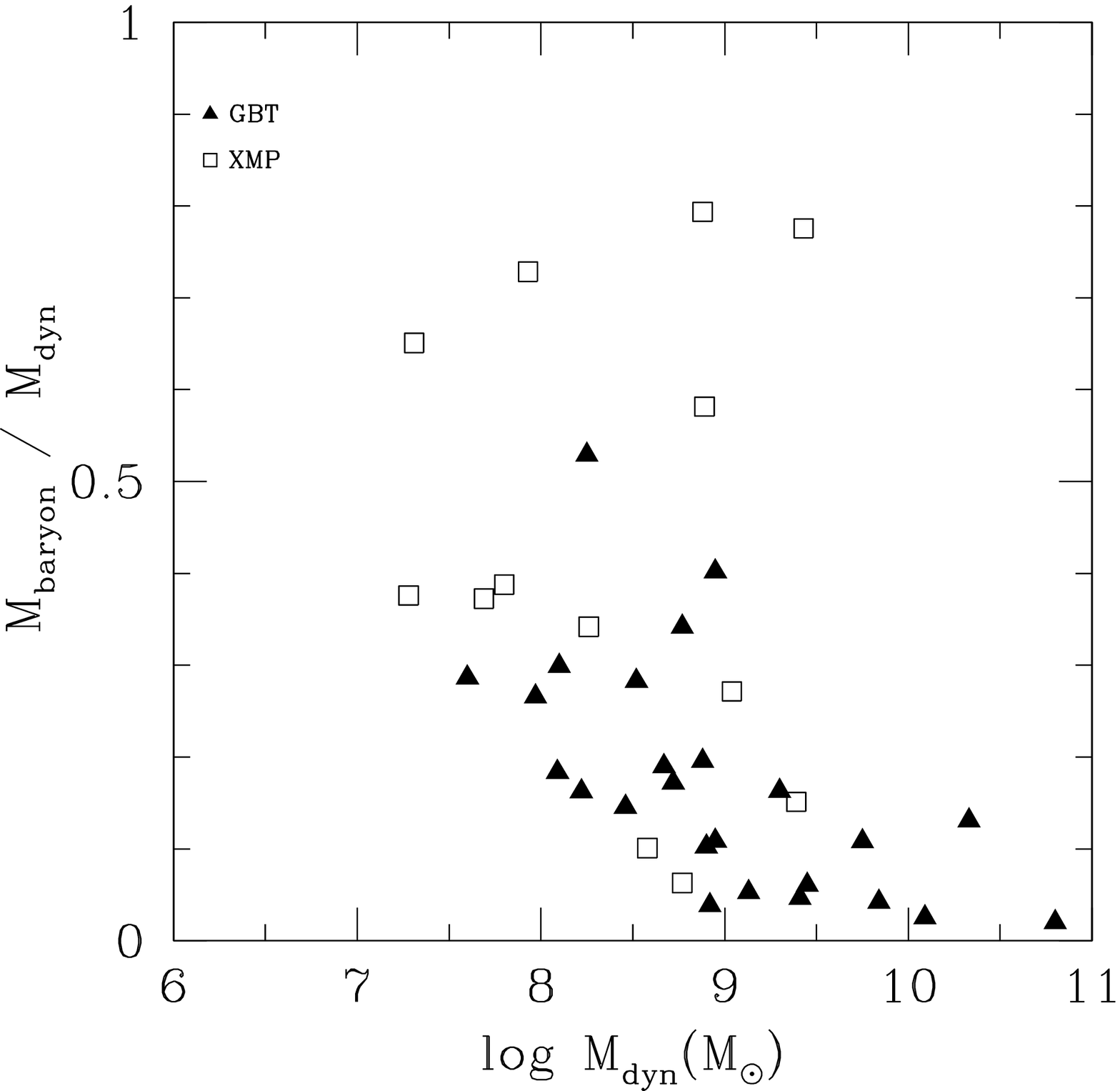}
\caption{Plot of the baryon to dynamical mass fraction versus the 
dynamical mass for the GBT and XMP samples (the SINGS and Engelbract samples do not have published dynamical masses).} 

\label{fig12}
\end{figure}

\subsection{Baryonic mass fraction}



Fig. \ref{fig12} shows the plot of the baryonic mass fraction 
$M_{baryon}$ versus the dynamical mass $M_{dyn}$ for the GBT and 
XMP samples. The Engelbracht and SINGS samples do not have readily available 
dynamical mass data.  
  No correlation is evident in the dynamical mass range 
beween 10$^{7}$ and 10$^{10.5}$~M$_\odot$. The baryonic fraction of the 
GBT and XMP objects varies from 0.05 to 0.80, with 
a median value of $\sim$0.2.
This is similar to the results of \citet{Bra15} who found  
for their large dwarf and non-dwarf SDSS galaxy sample
a median baryon to dynamical mass ratio of 0.15$\pm$0.18, although this 
agreement may be somewhat 
fortuitous as these authors calculate $M_{dyn}$ differently 
from us: 
they use the H {\sc i} velocity width at 20\% instead of 50\% of maximum intensity, and the radius of the H {\i} component is derived from a statistical 
relation between H {\sc i} mass and H {\sc i} radius. Our value 
is slightly smaller than the median baryon to dynamical mass ratio 
(within 3 optical scale lengths) 
of $\sim$0.3 obtained by \citet*{Lel14}, using H {\sc i} 
rotation curves derived from interferometric maps. These
 baryonic mass fractions 
derived from H {\sc i} rotation curves are likely more reliable. However,  
given that our dynamical masses are only approximate estimates from the 
H {\sc i} velocity widths, the relatively good agreement implies that 
our dynamical mass estimates are not too far off.



\begin{figure*}
  \centering
\hbox{
  \includegraphics[width=.396\linewidth]{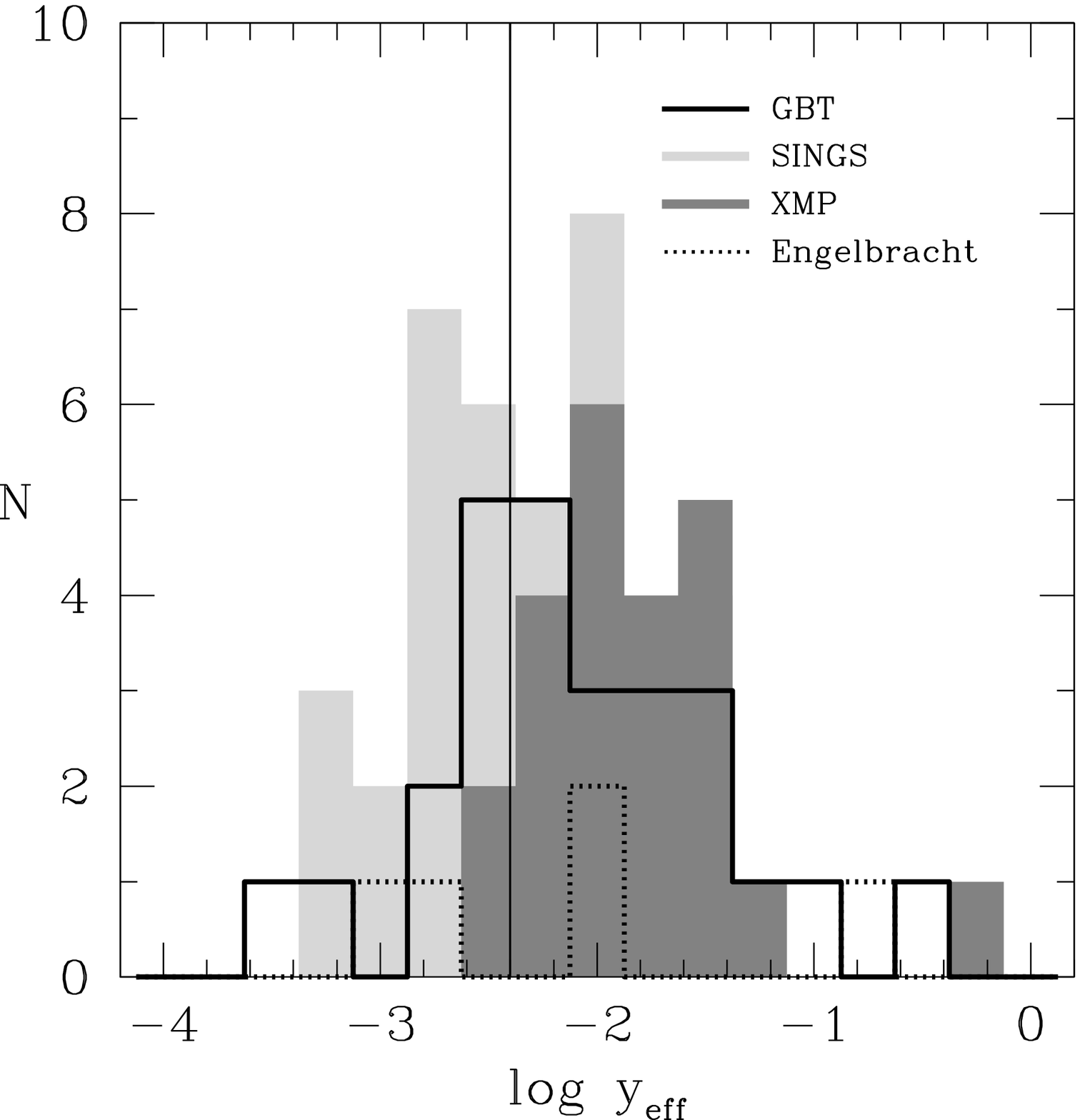}
	\hspace{0.3cm}\includegraphics[width=.4235\linewidth]{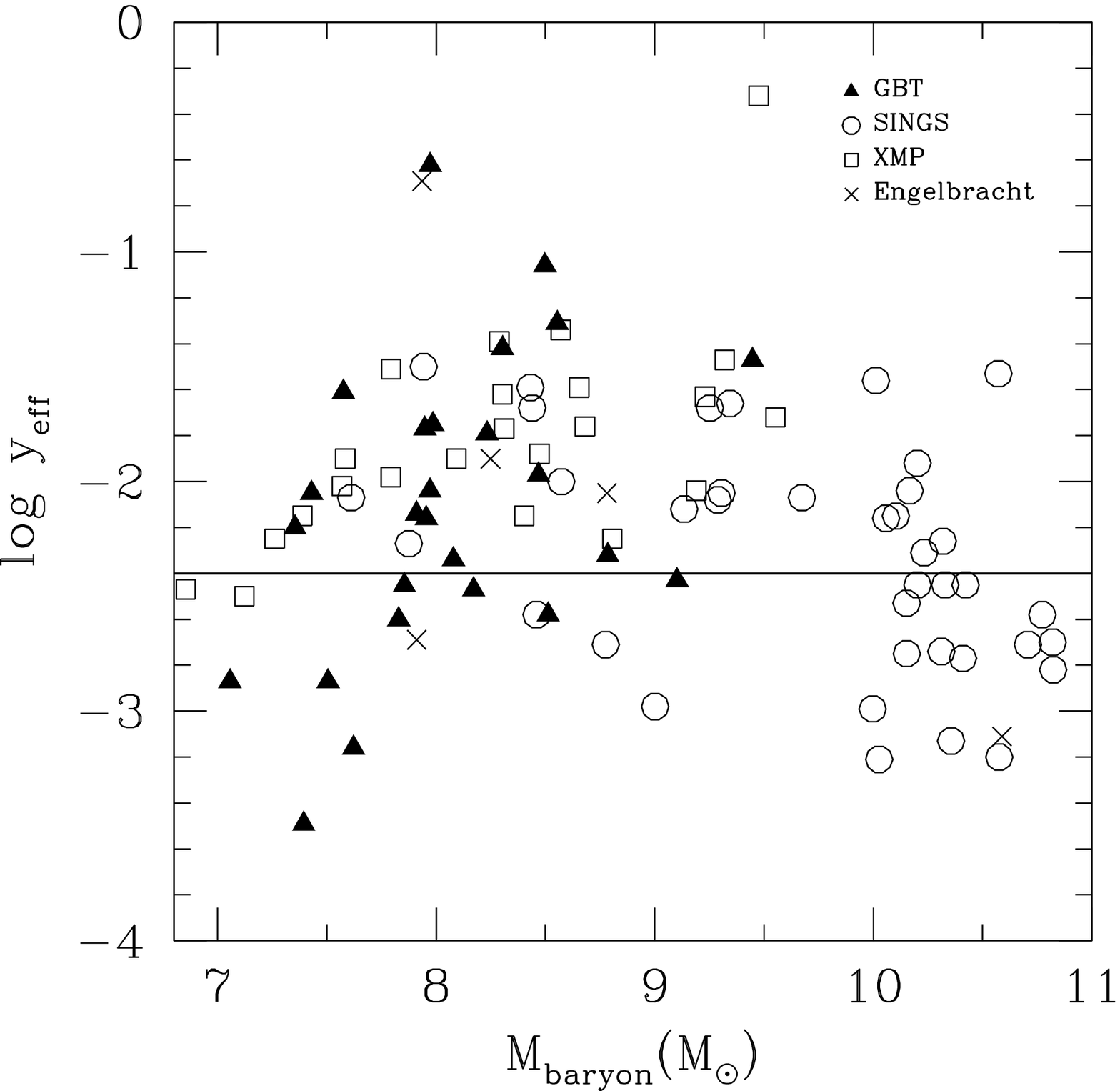}}
	\caption{(left) Histogram of the effective yield $y_{\rm eff}$ for the 
4 galaxy samples; (right) Plot of 
$y_{\rm eff}$ vs. $M_{baryon}$. The solid line shows the true yield for a closed-box model, log $y$ = $-$2.4 \citep{Dal07}.}
\label{fig13}
\end{figure*}

\subsection{Chemical Evolution}

Knowledge of both the gas mass and the (ionized) gas metallicity allows us 
to test chemical evolution models of galaxies. In particular, 
for galaxies that evolve without gas 
infall or wind outflow, i.e. that can be described by 
closed box chemical evolution models, it is 
 predicted that the effective stellar yield 
$y_{\rm eff}$
is a simple function of the metallicity $Z_{\R{gas}}$ (in units of mass fraction)
and the gas mass fraction 
$f_{\R{gas}}$.  

\begin{equation}
y_{\R{eff}} = \frac{Z_{\R{gas}}}{\R{ln}(1/f_{\R{gas}})} \, ,
\end{equation}

\noindent
where $Z_{\R{gas}}$ = 12 $\times$(O/H).

Here, we use the oxygen abundance as a proxy for the gas metallicity, and 
the factor 12 is the conversion from the number ratio O/H to oxygen 
mass fraction \citep{Gar02}.
If the galaxy truly evolves as a closed box, 
then the effective yield $y_{\rm eff}$ should be equal 
to the true yield $y_{\rm true}$ calculated from stellar evolution models.
These models give   
log $y_{\rm true}$ $\sim$-2.4 \citep{Dal07}.

In Fig. \ref{fig13} (left), we plot the histograms 
of $y_{\rm eff}$ for the four samples. The  $y_{\rm true}$ value is shown by the 
solid vertical line.
It is seen that while the  $y_{\rm true}$ line bissects the data, there are 
many galaxies with  $y_{\rm eff}$ inferior to or superior to $y_{\rm true}$. 
\citet{Edm90} has shown that 
$y_{\rm eff}$ would be lower than $y_{\rm true}$ if metals have been lost from the 
galaxy through supernova-driven outflows, 
or if the current gas in the galaxy has been diluted 
with inflows of metal-poor gas.    
But how do we understand objects with effective yields greater than 
the theoretical yield? As discussed by \citet{Fil13},  
this can be accounted for by either an underestimate of the true yield, or an overestimate of the effective yield, or both.
For example, the true yield can 
be underestimated if the stellar Initial Mass Function (IMF) depends  
on metallicity, for example if the IMF slope flattens with decreasing 
metallicity \citep[e.g. ][]{Bro04}. A top-heavy IMF would give a higher 
true yield. However, the 
spectral energy distributions of low-metallicity 
star-forming dwarf galaxies are well 
fitted with an IMF with a Salpeter slope, and do not show evidence for 
a top-heavy IMF \citep[e.g. ][]{Izo11}. 
It is more likely that the effective yields have been overestimated. 
In calculating it, we have assumed that 
the metallicity of the neutral gas is equal 
to that of the ionized gas, an assumption not likely to be true.
In fact, we can estimate 
the ratio of the ionized gas metallicity to that of the neutral gas 
metallicity as it is roughly equal to  
the ratio of the effective yield to the true yield \citep{Fil13}. 
Fig. \ref{fig13} (right) shows that, for the majority of the GBT sample,  
this ratio ranges from about 1 to $\sim$20. This is very similar to the range 
of values found by  \citet*{Thu05} in their UV absorption line 
studies of BCDs. They show   
that the metallicity of their neutral gas is systematically smaller 
than that of their ionized gas, by a factor varying between 
$\sim$20 for the higher-metallicity BCDs, and $\sim$1.5-2 for 
the lowest-metallicity BCDs, such as I Zw 18 or SBS~0335$-$052E. This metallicity 
difference is likely due to the fact that  
metals produced in the H {\sc ii} regions have not
had time to diffuse out and mix with the H {\sc i} gas in the outer regions.
Thus the H {\sc i} gas in the envelope of BCDs is relatively metal-free.
This is in agreement with the results of \citet{Fil13} who found the ratio of 
the metallicity of the ionized to that of the neutral gas to range between 
1 and 10 for their XMP objects.

In summary, the distribution of the effective yields on either side of 
the true yield in 
Fig. \ref{fig13} (left) can 
be understood as  
gas outflow and/or inflow of unenriched gas in the case of objects 
with $y_{\rm eff}$ $\leq$  $y_{\rm true}$, and a relatively metal-free  H {\sc i} 
envelope for objects with $y_{\rm eff}$ $\geq$  $y_{\rm true}$.   

Fig. \ref{fig13} (right) shows $y_{\rm eff}$ as a function of the baryonic 
mass of the galaxy. Both dwarf and non-dwarf galaxies 
display a large scatter on either side of the true yield line 
$\log$  $y_{\rm true}$ $\sim -2.4$ \citep{Dal07}. 
 There is no evident variation of the 
effective yield with the baryonic mass of the galaxy. 
This result appears to be at odds with those obtained 
\citet{Gar02} and \citet{Tre04}. Analyzing a sample of $\sim$40 nearby  
spiral and irregular galaxies, \citet{Gar02} found $y_{\rm eff}$
to be constant for galaxies with rotational velocities $v_{rot}$ larger than 
$\sim$100 km s$^{-1}$, but to decrease by a factor of $\sim$10-20 below that 
threshold.  
 \citet{Tre04}, using the spectroscopic data base of $\sim$53,000 SDSS star-forming galaxies at $z\sim$0.1, and using indirect estimates of the gas mass based on the H$\alpha$ luminosity, also found that  $y_{\rm eff}$ decreases with
decreasing baryonic galaxy mass. Both sets of authors attribute the decrease 
of the effective yield at low galaxy masses to galactic winds 
removing metals more efficiently from the shallower potential wells of 
dwarf galaxies \citep[see also ][]{Sil01}.  
However, \citet{San14,San15} have argued that 
the low metallicities of the XMD galaxies are an indicator of infall of 
pristine gas. This process would increase the effective yield of 
low-mass galaxies as compared to more massive galaxies.   
The real situation is likely described by both gas outflow and infall, 
so that the effective yields of the GBT dwarf galaxies do not 
decrease significantly as compared to the effective yields 
of more massive galaxies (like the SINGS galaxies).

\section{Summary and Conclusions}

New H {\sc i} observations with the  Green Bank Telescope (GBT) 
are presented for a sample of 29 extremely metal-deficient 
star-forming Blue Compact Dwarf (BCD) galaxies.  
 The BCDs were selected from the spectal database of Data Release 7 
of the Sloan Digital Sky Survey (SDSS) to have a well detected 
[O{\sc iii}]$\lambda$4363 line (for direct abundance determination) and an oxygen abundance   
12 + log O/H $\leq$ 7.6. Neutral hydrogen was detected in 28 galaxies, a 97\% detection rate. For each galaxy, we have derived ancillary data from the
SDSS optical spectrum such as 
oxygen abundance, star formation rate, and 
stellar mass.

Because of the narrow metallicity range of the GBT sample (the lower limit 
of 12 + log O/H is 7.35), we have also added published 
H {\sc i} and optical data for three  
complementary galaxy samples to extend the metallicity and mass range and 
study statistically how the H {\sc i} content of a galaxy varies with
various global galaxian properties.
We have found the following: 

1)  
The lowest-luminosity lowest-metallicity galaxies have the largest
neutral hydrogen mass to light ratios, following 
the relation $M$(H {\sc i})/$L_g$
$\propto$ $L_g^{-0.3}$, in good agreement with the dependence 
found in previous studies of galaxy samples with a smaller luminosity range.
Our derived mass-metallicity relation is also in good agreement with those 
derived by other authors.  
 
2)  Metal-deficient low-mass dwarf galaxies are gas-rich. The median gas mass fraction of the GBT sample is 0.94, while its 
mean gas mass fraction is 0.90$\pm$0.15. 
The vast majority of the GBT galaxies have more than 65\% of their baryonic mass in gaseous form.
The existence 
of a lower limit, also found for larger dwarf samples, puts stringent 
constraints on feedback mechanisms in low-mass galaxies as they should not remove all of the 
galaxy's atomic gas.

3) The H {\sc i} depletion time is independent
of metallicity or stellar mass. Although there is a large scatter 
about the median value of 3.4 Gyr, the relative constancy of the 
gas depletion time implies that external processes or feedback mechanisms that control the gas supply are important for regulating star formation in both low- and high-mass galaxies.  

4) The ratio of the baryonic mass to the dynamical mass varies over a wide range, from 0.05 to 0.80, with a median value of $\sim$0.2 and no dependence on the dynamical mass.

5) 
About 35\% of the GBT galaxies have an effective yield 
less than the true yield, which can be understood as the result of the loss  
of metals due to supernova-driven outflows, and/or the consequence 
of dilution by inflows of metal-poor gas. 
However, the remaining 65\% 
show an effective yield larger 
than the true yield. This can be understood if the metallicity of the neutral gas is lower than the metallicity of the ionized gas by a factor $\sim$1.5--20,
as UV absorption studies of BCDs also show.











\section{Acknowledgements}

T.X.T. thanks the support of NASA grant GO4-15084X. The SDSS is managed by the Astrophysical Research Consortium for the Participating Institutions. 
This research made use of the NASA/IPAC Extragalactic Database (NED) which is 
operated by the Jet Propulsion Laboratory, California Institute of Technology,
under contract with NASA.

\end{document}